\documentclass[oneside, reqno, 11pt]{amsart}
 \let\counterwithin\relax \usepackage{graphicx,amsfonts,amssymb,amsmath,amsthm,url,amscd} \usepackage{enumitem} \usepackage{color} \usepackage{chngcntr}
\usepackage[dvipsnames]{xcolor} \usepackage[normalem]{ulem} \usepackage{comment}
\usepackage{graphicx, amsmath, amssymb, amsfonts, amsthm, tikz, amscd,mathtools}
\usetikzlibrary{matrix,arrows,decorations.pathmorphing}
\usepackage[a4paper, margin=2.7cm]{geometry} \usepackage{verbatim}
\usepackage{calc}
\usepackage{dsfont}
\usepackage{multirow}
\usepackage{tikz}
\usepackage{changepage}
\usetikzlibrary{quantikz2}
\usepackage[linesnumbered, ruled, vlined, algosection]{algorithm2e}
\SetKwInput{KwComplexity}{Complexity}
\SetKw{Continue}{continue}
\SetKw{False}{false}
\SetKw{True}{true}

\usepackage{thmtools}
\makeatletter
\@ifundefined{newcounteralias}{}{%
	\renewcommand\thmt@autorefsetup{%
		\@xa\def\csname\thmt@envname autorefname\@xa\endcsname\@xa{\thmt@thmname}%
	}%
}
\makeatother
\makeatletter
\def\thmt@innercounters{equation,algocf}
\makeatother

\counterwithin{figure}{section}
\counterwithin{table}{section}

\usepackage{graphicx, amsmath, amssymb, amsfonts, amsthm, stmaryrd, tikz, amscd} \usepackage[all]{xy} \usetikzlibrary{matrix,arrows,decorations.pathmorphing}

\usepackage[hyperfootnotes=false, colorlinks, citecolor= RoyalBlue, urlcolor=blue, linkcolor=blue ]{hyperref}

\usepackage{caption}
\usepackage{tabularray}
\usepackage{multicol}

\usepackage{etoolbox} \patchcmd{\section}{\scshape}{\bfseries}{}{} \makeatletter \renewcommand{\@secnumfont}{\bfseries} \makeatother

\usepackage{placeins}

\let\Oldsection\section
\renewcommand{\section}{\FloatBarrier\Oldsection}

\makeatletter
\let\orgdescriptionlabel\descriptionlabel
\renewcommand*{\descriptionlabel}[1]{%
	\let\orglabel\label
	\let\label\@gobble
	\phantomsection
	\protected@edef\@currentlabel{#1}%
	\let\label\orglabel
	\orgdescriptionlabel{#1}%
}
\def\paragraph{
	\@startsection{paragraph}{4}
	\z@{.5\linespacing\@plus.7\linespacing}{-.5em}%
	{\normalfont\itshape}}
\makeatother

\theoremstyle{plain} 
\newtheorem{theorem}{Theorem}[section]
\newtheorem{problem}[theorem]{Problem} \newtheorem*{problem*}{Problem}

\theoremstyle{definition}

\theoremstyle{plain} 
\declaretheorem[name=Proposition, sibling=theorem]{prop}

\theoremstyle{remark}

\newtheorem{definition}[theorem]{Definition}
\newtheorem*{definition*}{Definition}

 \newtheorem*{example*}{Example}

\newtheorem{remark}[theorem]{Remark} \newtheorem*{remark*}{Remark}       

\newtheoremstyle{itplain} 
{6pt} 
{5pt\topsep} 
{\itshape} 
{} 
{\itshape} 
{.}  
{5pt plus 1pt minus 1pt} 
{} 

\theoremstyle{itplain} 
\newtheorem{lemma}[theorem]{Lemma}

\newtheorem*{lemma*}{Lemma}

\newtheorem{corollary}[theorem] {Corollary} \newtheorem*{corollary*} {Corollary}

\theoremstyle{remark} 

\newtheorem*{lemmatest*}{Lemma}

\usepackage{etoolbox} \patchcmd{\section}{\scshape}{\bfseries}{}{} \makeatletter \renewcommand{\@secnumfont}{\bfseries} \makeatother

\renewcommand{\Re}{\mathrm{Re}} \renewcommand{\Im}{\mathrm{Im}}

\renewcommand{\mod}{\ \mathrm{mod}\,}

\newlength{\faktorheight}

\numberwithin{equation}{section}     \DeclareMathOperator{\GL}{GL}

\DeclareMathOperator{\Nr}{Nr} \DeclareMathOperator{\Mult}{Mult} \DeclareMathOperator{\Span}{span}

\newcommand{\nequiv}{\not \equiv}

\newcommand{\sm}{\left(\begin{smallmatrix}} \newcommand{\esm}{\end{smallmatrix}\right)} \newcommand{\bpm}{\begin{pmatrix}} \newcommand{\ebpm}{\end{pmatrix}}

\newcommand{\lquotient}[2]{ \mathchoice {
		\text{\lower1ex\hbox{$#1$}\Big \backslash \raise01ex\hbox{$#2$}}%
	} {
		#1\,\backslash\,#2 } {
		#1\,\backslash\,#2 } {
		#1\,\backslash\,#2 } }

\newcommand{\rquotient}[2]{ \mathchoice {
		\text{\raise01ex\hbox{$#1$}\Big/\lower1ex\hbox{$#2$}}%
	} {
		#1\,/\,#2 } {
		#1\,/\,#2 } {
		#1\,/\,#2 } }

\newcommand{\lrquotient}[3]{ \mathchoice {
		\text{\lower1ex\hbox{$#1$}\Big \backslash \raise01ex\hbox{$#2$}\Big/\lower1ex\hbox{$#3$}}%
	} {
		#1\,\backslash\,#2\,/\,#3 } {
		#1\,\backslash\,#2\,/\,#3 } {
		#1\,\backslash\,#2\,/\,#3 } }

\author{Yuval Spiizer}
\address{Tel Aviv University, Israel}
\email{yuval.lahav@mail.huji.ac.il}

\author{Raphael S. Steiner}
\address{Computing Systems Lab, Huawei Zurich Research Center, Thurgauerstrasse 80, 8050 Zurich,
	CH}
\email{raphael.steiner@huawei.com}

\date{\today} 

\title[Long rational and Gau{\ss}ian integer multiplication]{Long rational and Gau{\ss}ian integer multiplication}

\begin{document}

	\begin{abstract}
	Long integer multiplication is a fundamental operation in computational algebra, heavily relying on Toom--Cook and related evaluation-interpolation algorithms. Conventional wisdom dictates that multiplication on the Gaussian integers $\mathbb{Z}[i]$ is reduced to standard rational integer multiplication, since $\mathbb{Z}[i]$ is a ring extension of $\mathbb{Z}$. We show that this approach is far from optimal and moreover inverting this established paradigm, that is reducing rational to Gaussian multiplication, leads to faster rational integer multiplication algorithms.
	
	We introduce a novel Toom--Cook framework that reduces rational long integer multiplication to Gaussian integer multiplication, yielding a strictly more efficient computational pipeline. By taking advantage of the special properties of the rational integer domain inside the Gaussian integers, we achieve a more favourable algebraic structure that reduces the overall arithmetic cost in every metric. To validate this theoretical shift, we provide a complete implementation of our algorithm. Benchmarks demonstrate that our approach outperforms the highly optimised GNU Multiple Precision Arithmetic Library (GMP) by up to 29\%, bridging a fundamental theoretical insight with significant practical improvements in high-performance computer algebra.
\end{abstract}
	\maketitle
	\setcounter{tocdepth}{1}
	\tableofcontents
	
	\section{Introduction}
Long integer multiplication algorithms are widely used in many applications such as high precision calculations (cf.\@ \cite{brooks1981dynamics, emmart2011high}), cryptography (cf.\@ \cite{ding2018high, srikanth2020enhancement, hao2023efficient, das2024area}), quantum computing (cf.\@ \cite{kepley2015quantum,elia2021loss}), and artificial intelligence (cf.\@ \cite{ahmed2020atm,kalvacherla2023approximate}). Broadly speaking, there are three classes of such algorithms:
\begin{enumerate}[label=(\roman*)]
    \item Textbook, with arithmetic complexity $O(n^2)$;
    \item Toom--Cook-$k$, with arithmetic complexity $O(n^{\log_k(2k-1)})$ with implied constant depending on $k$ \cite{karatsuba-multiplication,toom,cook}; and
    \item FFT-based, with arithmetic complexity $O(n \: {\rm polylog}(n))$ \cite{schonhage, gaudry2007gmp, furer2009faster,  harvey2016even, harvey2021integer, OAI:Integer-multiplication-below-n-log-n-September-23-2026}.
\end{enumerate}
Despite the superior asymptotic complexity of FFT-based algorithms compared to the other classes, they include greater implied constants to the point that some are impractical. For this reason, Toom--Cook-$k$ algorithms are preferred for a wide range of inputs and are widely used in many operating systems (cf.\@ \cite{gmp}). Hence, extensive research efforts have been directed towards improving Toom--Cook-$k$ algorithms. This includes
\begin{itemize}
    \item reducing the arithmetic complexity, see \cite{zuras1993squaring,bodrato,zanoni2009Toom,zanoni2010iterative,gu2021optimized,spiizer2022toom,nissim2024fault};
    \item reducing the communication cost, see \cite{bilardi2019complexity,destefani2020communicationoptimalparallelstandardkaratsuba,de2022brief,nissim2024fault,nissim2024minimizing};
    \item reducing the runtime cost on processors in different systems, see \cite{ding2018high,yang2021521,allam2023comparative,kalvacherla2023approximate,wang2023tcpm,das2024area};
    \item reducing the number of gates, see \cite{fan2010overlap, kepley2015quantum, lee2015subquadratic, kim2022toffoli, thirumoorthi2022optimized}; and
    \item reducing the number and depth of gates in quantum computing settings, see \cite{dutta2018quantum,larasati2021quantum,putranto2023space,wardhani2024high}.
\end{itemize}

Bodrato and Zanoni \cite{bodrato} introduced the Toom-graph technique accompanied with a heuristic to improve the number of arithmetic operations required in the interpolation phase of Toom--Cook. Zanoni \cite{zanoni2010iterative} further reduced the number of arithmetic operations in the evaluation phase by reusing intermediate results. Spiizer \cite{spiizer2022toom} utilised the sparse-decomposition technique \cite{beniamini} to reduce the leading coefficient of Toom--Cook's arithmetic by introducing lower-order monomials and increasing the memory footprint of the algorithm. Later, Nissim, Schwartz, and Spiizer \cite{nissim2024minimizing} utilised the alternative-basis technique \cite{karstadt} to minimise the communication cost while keeping a low arithmetic complexity. The alternative-basis technique reduces the coefficient of the leading monomial of the arithmetic complexity at the cost of introducing an additive term of $n\log_k(n)$. The sparse-decomposition technique further reduces the coefficient of the leading monomial at the cost of increasing the coefficients of non-leading monomials and increasing the asymptotic communication cost of the algorithm. In contrast to their original use, matrix multiplication, these techniques dramatically increase the number of divisions in Toom--Cook algorithms by numbers which are not a power of two, so-called non-trivial divisions.
%

\begin{table}[!htpb]
\centering
\begin{tblr}{
		colspec = {  Q[c,m] || Q[c,m] | Q[c,m] | Q[c,m] | Q[c,m] | Q[c,m] | Q[c,m] | Q[c,m] },
		row{1} = {font=\bfseries},
		rowhead = 1,
		rowfoot = 0,
	}
\SetCell[r=2]{m}{Algorithm} & \SetCell[r=2]{m}{Leading \\ monomial} & \SetCell[c=6]{c}{Leading coefficient} & & & & &\\ \hline 
        
        & &\cite{zuras1993squaring}&\cite{bodrato}& \cite{zanoni2010iterative}&\cite{nissim2024minimizing}&\cite{spiizer2022toom}& {[Here]} \\
        \hline \hline
        Textbook&$n^2$& $4$ & --- & --- & --- & --- & ---\\
            \hline
        Karatsuba &$n^{\log_2 3}$& $10$ &---&---& --- & --- & --- \\
            \hline
        Toom--Cook-$3$ &$n^{\log_3 5}$&$33.5$ & $23.5$ &$22.5$&$21.5$&$\boldsymbol{20.5}$&$\boldsymbol{20.5}$ \\
            \hline
        Toom--Cook-$4$ &$n^{\log_4 7}$&$59.67$ & $38.67$&$32.67$&$34$&$\boldsymbol{30}$& $\boldsymbol{30}$ \\
            \hline
        Toom--Cook-$5$ &$n^{\log_5 9}$& $82.75$ &$53.75$&$46.75$&$46.25$& $40.25$  & $\boldsymbol{33.75}$
\end{tblr}

\caption{Leading term of the arithmetic complexity, see \S\ref{sec:computational-model},  of various long rational integer multiplication algorithms, where the number of digits in each input is denoted by $n$. Our algorithm utilises the respective reduction-to-Gaussian Toom--Cook-$k$ together with the respective Gaussian Toom--Cook-$k$ algorithm for Gaussian integers as the long Gaussian integer multiplication algorithm.
} \label{table:main-time-complexity}
\end{table}

\begin{figure}
    \centering
    \includegraphics[width=0.95\linewidth]{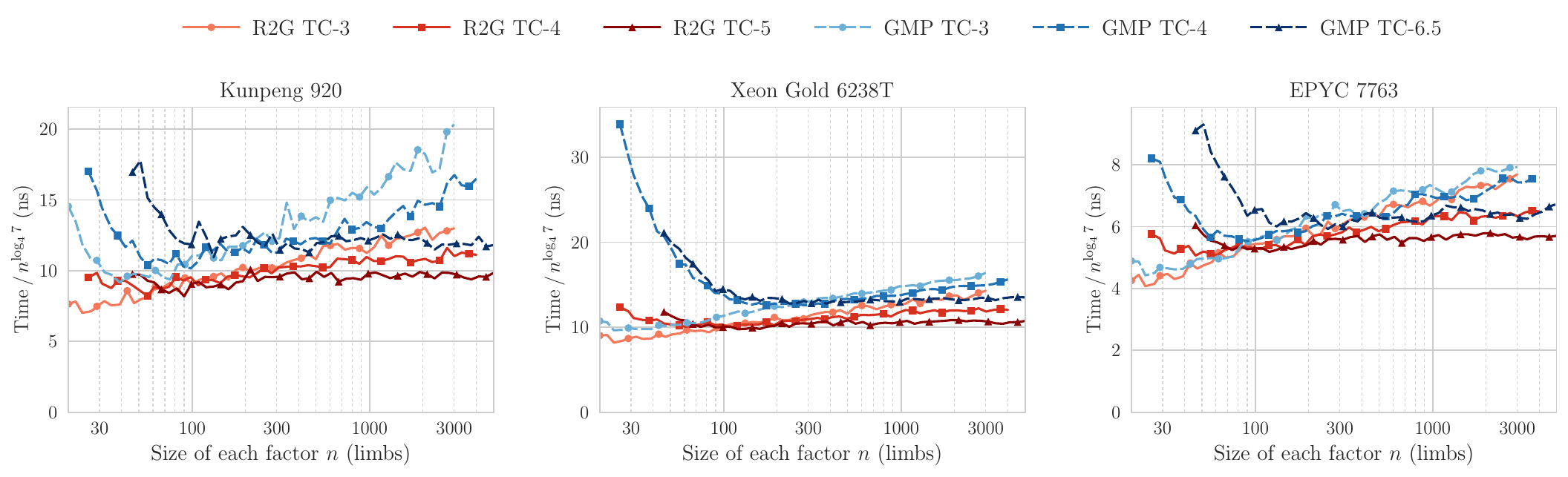}
    \caption{A comparison between our reduction multiplication algorithm and GMP 6.3.0 on different machines. Our reduction algorithms (R2G TC) are in shades of red, while GMP Toom--Cook algorithms (GMP TC) are in shades of blue. On the x-axis are the number of limbs $n$ (each limb is 64 bits) of each integer on a logarithmic scale. On the y-axis, we have the time in nano seconds (ns) per $n^{\log_4 7}$, that is the number of limbs $n$ raised to the $(\log_4 7)$-th power. Note that GMP TC-6.5 uses the balanced version when the inputs are balanced, which is the case here.}
    \label{fig:intro_shared_plot}
\end{figure}

Table~\ref{table:main-time-complexity} contains an overview of the arithmetic complexity of Toom--Cook-$k$ algorithms from the literature for small $k$ and Figure~\ref{fig:intro_shared_plot} compares our new multiplication algorithms against their counterpart in the GNU Multiple Precision Arithmetic Library (GMP)~\cite{gmp}.
Here and throughout, arithmetic complexity refers to rational integral arithmetic complexity, capturing the number of rational integer additions, subtractions, multiplications, and divisions of a fixed size. Thus, a complex addition of fixed size is considered twice as expensive as its rational counterpart. Likewise, a complex multiplication of fixed size is six-times more expensive, cf.\@ \S\ref{sec:computational-model}. Whilst this is the generally accepted way to compare multiplication algorithms, it does not take into account the respective costs of said operations in compute systems. Non-trivial divisions are an example of a particularly expensive operation. In runtime systems, non-trivial divisions can have a large impact on performance as they are around four times slower to compute than integer multiplications \cite{granlund1994division,chung2007asymmetric,Fog-InstructionTables} and over ten times slower than multiplication and division by powers of $2$~\cite{Fog-InstructionTables}. Thus, much effort has gone into reducing them~\cite{bodrato, zanoni2009Toom, larasati2021quantum} or eliminating them altogether~\cite{montgomery1985modular, chung2007asymmetric, ding2018high, gu2018division, gu2020novel} in Toom--Cook algorithms as well as replacing the costly division instructions with cheaper instructions through the use of modular arithmetic, $2$-adic numbers, and Hensel's lemma \cite{Davis1985, purdy-linear-division, jebelean-exact-div-93, brent2010modern, gmp}.

Bodrato and Zanoni \cite{bodrato} used the Toom-graph technique to obtain $2k-5$ non-trivial divisions in Toom--Cook-$k$ algorithms for $k=3,4,5$. They furthermore conjectured that every \emph{rational} integer Toom--Cook-$k$ algorithm requires at least $2k-5$ non-trivial divisions. This has been recently proven by Nissim--Schwartz--Spiizer \cite{Toom-Cook-divisions}. In Table \ref{table:main-arithmetic-operations}, we give an overview of the number of non-trivial divisions in Toom--Cook-$k$ algorithms from the literature for small $k$.

\begin{table}[htbp]
\centering
\begin{tblr}{
		colspec = {  Q[c,m] || Q[c,m] | Q[c,m] | Q[c,m] | Q[c,m] | Q[c,m] | Q[c,m] | Q[c,m] | Q[c,m] },
		row{1} = {font=\bfseries},
		rowhead = 1,
		rowfoot = 0,
	}
\SetCell[r=3]{m}{Algorithm} & \SetCell[c=8]{c}{Number of non-trivial divisions}& &&&&&&\\ \hline 
& \SetCell[c=6]{c}{Upper bound}& &&&&&\SetCell[c=2]{c}{Lower bound} &\\ \hline
        
        &\cite{zuras1993squaring}&\cite{bodrato}&\cite{zanoni2010iterative}&\cite{nissim2024minimizing}&\cite{spiizer2022toom}& {[Here]} & { $\mathbb{Z}$ \\ \cite{Toom-Cook-divisions} } & { $\mathbb{Z}[i]$ \\ {[Here]} } \\
        \hline \hline
        Textbook  & $\boldsymbol{0}$ & --- & --- & --- & ---& ---&---&---\\
            \hline
        Karatsuba & $\boldsymbol{0}$ &---&---& --- & ---& --- &---&--- \\
            \hline
        Toom--Cook-$3$ & $4$ & $1$ & $1$ & $2$ & $1$ & $\boldsymbol{0}$ &$1$&$0$\\
            \hline
        Toom--Cook-$4$ & $16$ & $3$ & $3$ & $5$ & $4$ & $\boldsymbol{1}$& $3$&$1$ \\
            \hline
        Toom--Cook-$5$ & $74$& $5$ & $5$ & $9$ & $7$ & $\boldsymbol{3}$ & $5$ &$3$
\end{tblr}
\caption{The maximum out of the number of non-trivial divisions in the rational (and Gaussian) base iteration(s). Our algorithm utilises the respective reduction-to-Gaussian Toom--Cook-$k$ together with the respective Gaussian Toom--Cook-$k$ algorithm for Gaussian integers as the long Gaussian integer multiplication algorithm.} \label{table:main-arithmetic-operations}
\end{table}

It is evident that our algorithm has fewer non-trivial divisions than the theoretical minimum for \emph{rational} integer Toom--Cook-$k$ algorithms. This is not a contradiction as our algorithm uses \emph{Gaussian} integer Toom--Cook-$k$ algorithms, thereby circumnavigating the condition of the theorem by Nissim, Schwartz, and Spiizer~\cite{Toom-Cook-divisions}.

Besides the number of non-trivial divisions, the magnitude of the divisors also plays a significant role as it leads to unnecessary information which is carried around. This is referred to as loss of precision \cite{elia2021loss}. This is most prominent in so-called \emph{division-free} Toom--Cook algorithms, which postpone all intermediate non-trivial divisions and replace them with a single terminal division. The first division-free Toom--Cook-$3$ algorithm was introduced by Gu and Li~\cite{gu2018division} in order to reduce the complexity in a quantum setting as the Toffoli depth and count are dominated by the non-trivial divisions in the interpolation phase, see also \cite{larasati2021quantum, putranto2023space}. Higher order division-free Toom--Cook-$k$ algorithms were presented by \cite{putranto2023space} and Wardhani--Putranto--Kim~\cite{wardhani2024high}, the latter also includes a quantum-circuit realisation. We also mention the implementations by Hao--Wang--Dang--Wang \cite{hao2023efficient} and Das--Jajodia~\cite{das2024area}. The terminal divisor in this method is determined by the \emph{division factor} of a Toom--Cook algorithm. The latter depends only on the evaluation points and is expected to grow rapidly with the order $k$ due to the increasing number of non-trivial divisions required for interpolation. A lower bound was provided by Nissim, Schwartz, and Spiizer \cite{nissim2025minimizing} including a new set of evaluation points that meet the lower bound for Toom--Cook-$4$ and $5$.
In Table~\ref{table:main-division-factor}, we present the division factor of our and various Toom--Cook algorithms, including the lower bounds from \cite{nissim2025minimizing} and our generalisation to Gaussian Toom--Cook. The division factor of our Toom--Cook reduction is smaller than the lower bound for rational Toom--Cook algorithms provided by \cite{nissim2025minimizing} because we sidestep their theorem by using Gaussian integers.

\begin{table}[htpb]
\centering
\begin{tblr}{
		colspec = {  Q[c,m] || Q[c,m] | Q[c,m] | Q[c,m] | Q[c,m] | Q[c,m]},
		row{1} = {font=\bfseries},
		rowhead = 1,
		rowfoot = 0,
	}
\SetCell[r=3]{m}{Algorithm} & \SetCell[c=5]{c}{Division factor}& &&&\\ \hline 
& \SetCell[c=3]{c}{Upper bound} & & & \SetCell[c=2]{c}{Lower bound} & \\ \hline
        
        & {\cite{zuras1993squaring},\cite{bodrato},\cite{zanoni2010iterative}, \\ \cite{nissim2024minimizing},\cite{spiizer2022toom} }& \cite{nissim2025minimizing}&{[Here]}& {$\mathbb{Z}$ \\ \cite{nissim2025minimizing} }& { $\mathbb{Z}[i]$ \\ {[Here]} }\\
        \hline \hline
        Textbook  & $\boldsymbol{1}$ & --- & --- & --- & --- \\
            \hline
        Karatsuba & $\boldsymbol{1}$ &---&---& --- & --- \\
            \hline
        Toom--Cook-$3$ & $3$  & $3$ & $\boldsymbol{1}$ & $3$ &$1$\\
            \hline
        Toom--Cook-$4$ & $45$ & $15$ & $\boldsymbol{5}$ & $15$& $5$ \\
            \hline
        Toom--Cook-$5$ & $2835$ & $315$ & $\boldsymbol{5}$ & $315$ & $5$
\end{tblr}
\caption{The least common multiple of the division factor(s) in the rational (and Gaussian) base iteration(s). Our algorithm utilises the respective reduction-to-Gaussian Toom--Cook-$k$ together with the respective Gaussian Toom--Cook-$k$ algorithm for Gaussian integers as the long Gaussian integer multiplication algorithm.}
\label{table:main-division-factor}
\end{table}

\subsection{Gaussian integer multiplication}

Unlike long rational integer multiplication, long Gaussian integer multiplication has been addressed rarely as an independent problem, but rather as a sub-problem of long rational integer multiplication. Peter Ungar suggested an algorithm that computes the product of two Gaussian integers using three long integer multiplications (according to {\cite[p.~647]{knuth1981semi}}). Earlier, Winograd had proven that at least three long integer multiplications are needed \cite{winograd1971multiplication}. More generally, long Gaussian integer arithmetic algorithms have often been based on long integer arithmetic algorithms (cf.\@ \cite{caviness1973lehmer,caviness1976algorithms,gilbert1984arithmetic,gilbert1996division,koval2016algorithm}).

\begin{table}[htpb]
	\centering
	\begin{tblr}{
			colspec = { Q[c,m] || Q[c,m] | Q[c,m] | Q[c,m] || Q[c,m] | Q[c,m] },
			row{1,2} = {font=\bfseries},
			rowhead = 1,
			rowfoot = 0,
		}
		\SetCell[r=3]{m}{Algorithm} & \SetCell[c=3]{c}{Arithmetic complexity} && & \SetCell[c=2, r=2]{h}{Non-trivial divisions}& \\
		\hline
		& \SetCell[r=2]{m}{Leading \\ monomial} & \SetCell[c=2]{c}{Leading coefficient} & & & \\
		\hline
		&& {$3$ long \\ $\mathbb{Z}$ mult.} & [Here] & {$3$ long \\ $\mathbb{Z}$ mult.} & [Here] \\
		\hline \hline
		Toom--Cook-$3$ & $n^{\log_3 5}$ & $61.5$ & $\boldsymbol{41}$  & $1$ & $\boldsymbol{0}$\\
		\hline
		Toom--Cook-$4$ & $n^{\log_4 7}$ & $90$ & $ \boldsymbol{60}$  & $4$ & $\boldsymbol{1}$ \\
		\hline
		Toom--Cook-$5$ & $n^{\log_5 9}$ & $120.75$ & $\boldsymbol{67.5}$  & $7$ & $\boldsymbol{3}$
	\end{tblr}
	\caption{Number of non-trivial divisions in the base iteration and leading term of the arithmetic complexity of our long Gaussian integer multiplication algorithms juxtaposed against using 3 long rational multiplications using the algorithm from \cite{spiizer2022toom}. The number of digits of the real and imaginary part of each input is $n$.
	} \label{table:complex-time-complexity}
\end{table}

\begin{figure}
	\centering
	\includegraphics[width=0.95\linewidth]{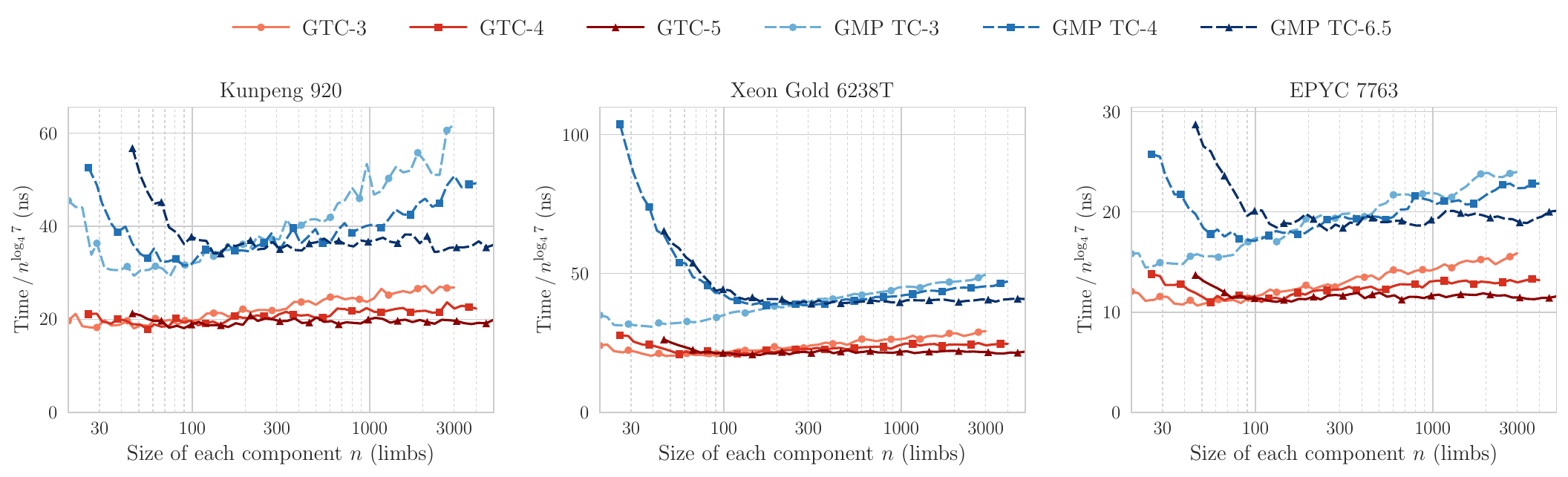}
	\caption{A comparison between our Gaussian Toom--Cook multiplication algorithm and computing a Gaussian product using three rational multiplications with GMP 6.3.0 on different machines. Our Gaussian Toom--Cook algorithms (GTC) are in shades of red, while GMP Toom--Cook algorithms (GMP TC) are in shades of blue. On the x-axis are the number of limbs $n$ (each limb is 64 bits) of each integer on a logarithmic scale. On the y-axis, we have the time in nano seconds (ns) per $n^{\log_4 7}$, that is the number of limbs $n$ raised to the $(\log_4 7)$-th power. Note that GMP TC-6.5 uses the balanced version when the inputs are balanced, which is the case here.}
	\label{fig:intro_Gauss_shared_plot}
\end{figure}

For the long Gaussian integer multiplication algorithm, we use Toom--Cook algorithms with interpolation points in $\mathbb{Q}(i) \cup \{\infty\}$. There, the interpolation points can be more densely packed and enjoy additional symmetries. However, this comes at the cost of more expensive additions, multiplications, and divisions. In Table \ref{table:complex-time-complexity}, we summarise the number of non-trivial divisions and upper bounds on the arithmetic complexity of Gaussian Toom--Cook-$k$ algorithms that we are able to achieve and compare them against using three long rational Toom--Cook-$k$ multiplications. In Figure~\ref{fig:intro_Gauss_shared_plot}, we show how these theoretical gains turn into practical gains on real hardware.

In our algorithms, there are three main features of the Gaussian integers that we exploit:
\begin{enumerate}[label=(\roman*)]
	\item multiplication by $\pm i$ is essentially free as it only entails a swap of real and imaginary part as well as a sign change;
	\item the rational prime $3$ remains prime in the Gaussian integers; and
	\item the Gaussian integers contain the $4$-th roots of unity and the $8$-th roots of $16$. \label{item:feature-roots}
\end{enumerate}

We note that the first two features have recently been used in the related context of computing two-dimensional convolutions \cite{GaussianWinograd}. In their and our work, the second feature is quintessential as it allows us to avoid a division by a multiple of $3$ in the Gaussian Toom--Cook-$3$ algorithm, whereas it is not possible to do so for a rational Toom--Cook-$3$ algorithm, see Proposition \ref{prop:division_by_p} and Remark \ref{rem:single-division-p}. Instead, the first necessary non-trivial division for Gaussian Toom--Cook-$k$ occurs for $k=4$, where a division by a multiple of $2+i$ must occur. More generally, we extend the main theorem of \cite{Toom-Cook-divisions} and show that the number of non-trivial divisions grows linearly with $k$.

\begin{theorem} \label{thm:main-divisions}
	Every Gaussian Toom--Cook-$k$ algorithm with $k \ge 2$ contains at least $2k-4$ divisions (by a non-unit) and $2k-7$ divisions by a Gaussian integer which is not a power of $1+i$ times a unit.
\end{theorem}

The final feature, that is the use of roots of unity \ref{item:feature-roots}, has been suggested as early as 1980 by Winograd \cite{winograd1980arithmetic} and has found its way into papers on multiplication in polynomial rings over finite fields \cite{bodrato2007notes}. The feature enables fast and efficient evaluation and interpolation schemes based on the discrete fast Fourier transform \cite{cooley1965algorithm, duhamel1990fast, johnson2006modified} when $k=4, 5$.

\subsection{Optimising arithmetic cost}

Evaluating the linear transformations inherent to the evaluation and interpolation phases of Toom--Cook algorithms can be framed as a search for a shortest linear straight-line program. The Toom-graph technique \cite{bodrato} does exactly that, but only for the interpolation phase. While highly effective, the shortest-path formulation of the Toom-graph inherently restricts the ability to reuse already computed intermediate steps and is limited to basis-preserving transformations. One can generalise this approach by modelling the computation of these linear forms as a minimum-weight Steiner-tree problem within a directed hypergraph \cite{steiner-tree-problems,gallo-hypergraphs}.

Steiner trees refer to a family of optimisation problems where the primary goal is to find a minimum-weight network connecting a given set of terminal nodes \cite{steiner-tree-problems}. While traditionally applied in Euclidean and rectilinear spaces for VLSI routing and network topology design \cite{steiner-tree-problems}, the generalisation of this problem to directed hypergraphs provides a powerful framework for modelling computational sequences \cite{gallo-hypergraphs}. In algebraic complexity, evaluating a set of linear forms can be formally mapped to finding a minimum-weight Steiner tree within a directed hypergraph, where hyperedges represent elementary arithmetic operations and nodes represent intermediate computed values \cite{gallo-hypergraphs}. 
This hypergraph formulation is mathematically equivalent to the Shortest Linear Program (SLP) problem, an NP-hard optimisation task \cite{BMP13}. Historically, SLP methodologies have seen extensive use in cryptography, where heuristic algorithms such as those leveraging target-distance prefixes are deployed to minimise the logic gate count of purely linear subcircuits over $\mathbb{F}_2$, like cryptographic S-boxes \cite{BMP13}. Beyond heuristic approaches, researchers have successfully synthesised exact shortest linear straight-line programs for Boolean applications by mapping the Steiner-tree search space to satisfiability problems, allowing for the use of automated SAT solvers \cite{FS10}. While these tools have driven significant advancements in boolean logic, adapting the Steiner-tree and SLP methodologies to minimise arithmetic operations over algebraic rings represents a novel transition from logic synthesis to exact arithmetic algorithm design.

To systematically discover optimal evaluation and interpolation sequences for our reduction-to-Gaussian and Gaussian Toom--Cook algorithms, we model the matrix inversion as a weighted shortest linear straight-line program problem over the Gaussian integers. Because this optimisation is inherently NP-hard \cite{BMP13}, we deploy a hybrid heuristic search architecture that combines a level-synchronous beam search (ranking states using a potential function akin to weighted $A^*$ \cite{Poh70}), 
and a ruin-and-recreate local search \cite{SSSWD00} over incumbent programs. The framework systematically explores write-once elementary row operations, where each step is strictly priced according to the computational model \S\ref{sec:computational-model}: unit multiplications are free, simple additions and powers of two cost 2, while general Gaussian scalars incur higher costs between 4 and 8. To aggressively prune the search space and maintain tractability, the engine leverages structural bounds, such as utilising the rank defect of the matrix pool at the prime $1+i$ to accurately bound the necessary divisions remaining in the sequence. 

The efficiency of this automated search is further driven by highly calibrated heuristics and algebraic pattern recognition. Rather than relying purely on incremental scoring, the framework builds structural patterns upfront by identifying FFT-like butterfly openings \cite{cooley1965algorithm} for node sets closed under negation and conjugation, effectively halving support sizes early in the computation. Furthermore, the heuristic potential is rigorously calibrated to real execution costs via prefix regressions \cite{TDR11}, while column-distinctness constraints ensure that shared operations are priced correctly without false overlaps. The heuristic also relies on smart Gaussian elimination completion, similar to the heuristic used for the Toom-graph technique \cite{bodrato}, but goes much further, like looking for FFT-like butterflies \cite{cooley1965algorithm}. This robust search methodology successfully synthesises highly efficient programs that consistently equal or surpass manually derived sequences.

\subsection{Our contributions}

Our main innovation is the incorporation of long \emph{Gaussian} integer multiplication into long rational integer multiplication, structurally reducing the overall arithmetic cost. This reverses the established problem hierarchy in computational algebra. Instead of computing long Gaussian integer multiplication using long rational integer multiplication algorithms, we compute long rational integer multiplication using a long Gaussian integer multiplication algorithm. We achieve this through a novel Toom--Cook-like reduction between the two domains. 

Specifically, our contributions are as follows:

\begin{itemize}
    \item \textbf{An inverted reduction hierarchy:} We utilise the automorphism properties of the complex conjugate to transform a set of multiplications in the standard rational Toom--Cook algorithm into a set of half as many complex multiplications. Moreover, our reduction is so efficient that it fully recovers the doubled cost of Gaussian addition and Gaussian-rational multiplication/division. By migrating the evaluation and interpolation sequences to $\mathbb{Q}(i)$, we can select interpolation points that are more densely packed and enjoy additional algebraic symmetries not present in the real domain. In Table~\ref{table:main-time-complexity}, we summarise the arithmetic complexity of our Toom--Cook reduction and compare it to previous results. Our model is based on a na\"ive implementation of Gaussian operations and does not allow any shortcuts (see \S \ref{sec:computational-model}).
    
    \item \textbf{Optimised Gaussian Toom--Cook algorithms:} While operations over $\mathbb{Q}(i)$ inherently introduce more computationally expensive additions, scalar multiplications, and divisions, we mitigate these overheads through targeted structural optimisations. We design and deploy DFT-like subroutines to reduce the arithmetic cost of the Gaussian Toom--Cook interpolation phase, and we utilise Steiner trees to optimise the evaluation phase. In Table~\ref{table:complex-time-complexity}, we summarise the number of non-trivial divisions and upper bounds on the arithmetic complexity of the Gaussian Toom--Cook-$k$ algorithms we achieve, comparing them against the cost of using three long rational Toom--Cook-$k$ multiplications.
    
    \item \textbf{Circumventing rational lower bounds:} Even though the asymptotic complexity of our algorithm matches that of the rational Toom--Cook-$k$ algorithm, our approach structurally requires fewer non-trivial divisions and yields a smaller division factor. Remarkably, the division factor and the number of non-trivial divisions of our algorithms fall strictly below the established lower bounds for rational Toom--Cook algorithms \cite{nissim2025minimizing,Toom-Cook-divisions}. Furthermore, we establish analogous lower bounds for our new Gaussian framework, which our algorithms strictly meet for $k \le 5$. In Table~\ref{table:main-arithmetic-operations}, we compare the number of non-trivial divisions and these lower bounds against previous algorithms. In Table~\ref{table:main-division-factor}, we provide a similar comparison for the division factor.
 
    \item \textbf{Steiner-tree optimisation for the evaluation phase:} To further improve the Gaussian Toom--Cook algorithm, we systematically optimise the evaluation phase using Steiner trees. By modelling the polynomial evaluation at our chosen complex interpolation points as a graph-theoretic network, we are able to identify near-optimal sub-expression sharing and minimal addition chains. This algorithmic approach strictly minimises the number of linear arithmetic operations required to apply the evaluation matrix, effectively neutralising the overhead of computing over $\mathbb{Q}(i)$ and contributing directly to our empirical speed-ups.

    \item \textbf{Escape local minimum using AI:} In the space of sequences of operations that compute the linear parts of the algorithm, there are many local minima. A traditional approach to escape a local minimum is to introduce random steps to the search. Also, knowledge of a better solution can help the search obtain a better result. We used by-hand improvements, and later used AI to find small improvements to the local minimum the search algorithm found, and then fed them back to the search algorithm. From a human perspective, the improvements that we found are close to the solution the search algorithm found, but they are far away from the perspective of the heuristic of the algorithm. This iterative mode of search-small improvement-search-small improvement, allowed us to reduce the inversion sequence of Gaussian Toom--Cook-$5$ from an arithmetic cost of $92$ to merely $72$.
    
    \item \textbf{Empirical state-of-the-art:} Theoretical improvements in long rational integer multiplication algorithms often mask large constant factors that render them impractical. To demonstrate the concrete impact of our algebraic shift, we implemented the algorithm and obtained real-world benchmarks (see Figure~\ref{fig:intro_shared_plot}). Our implementation structurally translates into a highly efficient computational pipeline, outperforming the heavily optimised GNU Multiple Precision Arithmetic Library (GMP) by up to 29\%, establishing a new practical standard for integer multiplication in this regime.

\end{itemize}

\subsection{Overview}

In \S \ref{sec:preliminaries}, we define the cost model precisely. In \S \ref{sec:toom-cook-description}, we introduce the rational-to-Gaussian reduction and Gaussian Toom--Cook schemes and bound their arithmetic complexities based on the arithmetic complexities of the evaluation and interpolation matrices. In \S \ref{sec:Steiner-tree}, we describe how Steiner trees can be used to optimise the evaluation and interpolation phases. In \S \ref{sec:alg-details}, we provide the details on the evaluation points used in our Toom--Cook schemes as well as efficient evaluation and interpolation sequences. In the sections \S\ref{sec:division-gaussian-toom-cook} and \S\ref{sec:division-factor}, we prove the lower bounds on divisions, Theorem~\ref{thm:main-divisions} and Table~\ref{table:main-division-factor}. In \S \ref{sec:bench}, we provide the empirical evaluation of our algorithms. In the final section, \S\ref{sec:discussion}, we discuss open problems.
	\subsection{Acknowledgements}
We would like to thank Renzo Andri and Yoav Moran for valuable discussions on this and surrounding topics. We also thank the anonymous reviewers for valuable comments and pointing us to division schemes based on Hensel's lemma.

\subsection{Funding}
This project has received funding from the European Research Council (ERC) under the European Union’s Horizon 2020 research and innovation programme (grant agreement No 818252, 101113120, 101138056). This project has received funding from the European Research Council under the European Union’s Horizon 2020 research and innovation programme (grant agreement No 101113120, 101138056). Research was supported by grant No. 1354/23 of the Israel Science Foundation (founded by the Israel Academy of Sciences and Humanities). This research has been supported by the Science Accelerator and by the Frontiers in Science initiative of the Ministry of Innovation, Science and Technology.

\subsection{Disclosure of AI use}
The authors of this paper were assisted by AI in the high-performance implementation of our algorithms. Likewise, AI was used in generating the optimisation framework around Steiner trees to find highly efficient evaluation and interpolation sequences. AI was further used for drafting and copy-editing paragraphs as well as generating figures for the paper. The authors retain full accountability for the accuracy and integrity of the final text and results.
	\section{Preliminaries}
\label{sec:preliminaries}
\subsection{Notations}

To clearly differentiate between concepts relating to the rational integers $\mathbb{Z}$ and Gaussian integers $\mathbb{Z}[i]$, we will use the adjectives `rational' and `Gaussian'. For example, $2$ is a rational prime, but not a Gaussian prime, and $(5-2i) \cdot 3$ is a multiplication of a Gaussian integer and a rational integer, but can also be regarded as a product of two Gaussian integers.




\subsection{Computational model} \label{sec:computational-model}

In this section, we describe the cost model, that is the arithmetic complexity, of elementary arithmetic operations. In what follows, a term or expression is referred to as \emph{fixed} if it is predetermined, say by the algorithm, or uniformly bounded. We do not make a distinction here as to how large the latter bound is.

\subsubsection{Operations with rational integers}
For the addition of two $n$-digit rational integers and the multiplication/division of an $n$-digit rational integer with/by a fixed rational integer, we charge a cost of $n$. As an exception to this rule, we charge nothing for the multiplication/division by $\pm 1$. For the multiplication/division of two fixed rational integers, we charge a cost of $1$. For the concatenation of two rational integers, we charge nothing. Likewise, for splitting a rational integer into a fixed number of pieces of (almost) equal size, we also charge nothing.


\subsubsection{Operations with Gaussian integers}
We say that a Gaussian integer $a+ib \in \mathbb{Z}[i]$ with $a,b \in \mathbb{Z}$ is an \emph{$n$-digit} Gaussian integer if and only if $a$ and $b$ are $n$-digit rational integers.

We align our cost model for Gaussian operations with the cost model for rational operations in the sense that the costs we charge for a Gaussian operation are the sum of the costs for a na\"ive implementation using rational operations on real and imaginary part.

For the addition of two $n$-digit Gaussian integers and the multiplication/division of an $n$-digit Gaussian integer with/by a fixed rational integer, we charge a cost of $2n$. As an exception to this rule, we charge nothing for the multiplication/division by $\pm 1$.

For the multiplication of an $n$-digit Gaussian integer with a fixed Gaussian integer, we charge $6n$ as this can be realised as four rational multiplications of an $n$-digit rational integer with a fixed rational integer and two additions of $n$-digit rational integers. As an exception to this rule, we do not charge anything for multiplications by $\pm i$. The reason for this exception is that the latter operation only entails a swap of the real and imaginary part and a rational sign change. For the multiplication of two fixed Gaussian integers, we charge a cost of $6$. For the division of an $n$-digit Gaussian integer by a fixed Gaussian integer, we charge $8n$ as this can be realised as a Gaussian multiplication and a rational division. For the division of two fixed Gaussian integers, we charge a cost of $8$. We again do not charge for concatenation and splitting into a fixed number of pieces of (almost) equal size of Gaussian integers.

We note that several Gaussian multiplications and divisions can be implemented with reduced costs. A multiplication by $1 \pm i$ can be realised with a cost of $2n$ instead of $6n$ by utilising Gaussian multiplication by a unit and Gaussian addition instead. Likewise, a division by $a(1 \pm i)$, with $a \in \mathbb{Z}$ fixed, costs $4n$ instead of $8n$ as this can be achieved by multiplication by $1 \mp i$ and division by $2a$. Finally, multiplications by $1+ai$, with $a \in \mathbb{Z}$ fixed, can be realised with cost of $4n$ instead of $6n$ and divisions by $1+ai$ can be realised with a cost of $6n$ instead of $8n$.

\subsubsection{Non-trivial divisions and the division factor}
We say that a division by an integer that is not a power of 2 is a non-trivial division. We count the number of non-trivial divisions in the base case of the algorithm as it predicts the number of non-trivial divisions in the entire algorithm. 

The division factor is recognised as the `depth' of the non-trivial divisions. Namely, the division factor $\alpha$ of a matrix $A$ is the minimal integer (or the Gaussian integer with minimal norm) such that there is $k\in \mathbb{Z}_{\ge 0}$ such that $2^k\cdot\alpha\cdot A$ has only (Gaussian) integer entries. 


	
	\section{Toom--Cook}
\label{sec:toom-cook-description}

Toom--Cook algorithms~\cite{toom, cook-phd-thesis} are long integer multiplication algorithms that are based on polynomial multiplication, using evaluation and interpolation. Here, we use an altered Toom--Cook scheme to reduce rational multiplication to Gaussian multiplication. We also describe the Toom--Cook algorithm in the Gaussian integers. The algorithms here are described in a projective way, that is, projective evaluation points $(x,y) \neq (0,0)$ here correspond to the affine evaluation points $x/y$ if $y \neq 0$ and the point $\infty$ if $y = 0$.

\subsection{Rational-to-Gaussian Toom--Cook reduction}
\label{sec:TC-reduction}

\subsubsection{Reduction scheme}
\label{sec:reduction-scheme}

Let $k\in \mathbb{Z}_{>0}$ and $T\subset \mathbb{Z}[i]^2\backslash \{(0,0)\}$ be a set of $2k-1$ pairs of Gaussian integers that project onto $2k-1$ distinct points on the projective line $P^{1}(\mathbb{Q}(i))$, that is, they are not scalar multiples of each other over $\mathbb{Q}(i)^{\times}$. We further assume that we can split $T = R \sqcup S \sqcup \overline{S}$ into rational points $R\subset \mathbb{Z}^2\backslash \{(0,0)\}$ and purely Gaussian points $S\subset \mathbb{Z}[i]^2\backslash \{(0,0)\}$ and their complex conjugates $\overline{S}$. The former meaning that their projections to $P^1(\mathbb{Q}(i))$ are in $P^1(\mathbb{Q})$ and the latter meaning that they are in $P^1(\mathbb{Q}(i)) \backslash P^1(\mathbb{Q})$.

Let $a$ and $b$ be two rational integers. To compute the product $a \cdot b$, we do the following:
\begin{enumerate}[label=(\Roman*)]
    \item Split $a$ and $b$ into $k$ parts each, denoted $[a]_j$, $[b]_j$, $j=0,\dots, k-1$, respectively, such that there is $M\in \mathbb{Z}_{> 0}$ with
    \begin{align}
        a=\sum_{j=0}^{k-1}[a]_j M^j, && b=\sum_{j=0}^{k-1}[b]_j M^j.
    \end{align}
    \item Define the polynomials $P_1, P_2 \in \mathbb{Z}[X,Y] \subset \mathbb{Z}[i][X,Y]$ as
    \begin{align}
            P_1(X,Y)=\sum_{j=0}^{k-1}[a]_jX^j Y^{k-1-j}, &&
            P_2(X,Y)=\sum_{j=0}^{k-1}[b]_jX^j Y^{k-1-j},
    \end{align}
    such that $P_1(M,1)=a$ and $P_2(M,1)=b$.
    \item For every $(x,y)\in R \sqcup S$, compute the pair $(P_1(x,y),P_2(x,y))$. \label{enum:Reduction_Toom_Cook_evaluation}
    \item For every $(x,y)\in S$, compute the product $Q(x,y)=P_1(x,y) \cdot P_2(x,y)$ using a Gaussian integer multiplication algorithm.
    \label{enum:Reduction_Toom_Cook_recursive_gauss_multiplication}
    \item For every $(x,y)\in R$, compute the product $Q(x,y)=P_1(x,y) \cdot P_2(x,y)$. If the size of either $P_1(x,y)$ or $P_2(x,y)$ is too big, compute the product recursively using the same procedure. \label{enum:Reduction_Toom_Cook_recursive_multiplication}
    \item Use interpolation to retrieve the coefficients of $Q = P_1 \cdot P_2$ from $Q(x,y)$, $(x,y)\in R \sqcup S$.
    \label{enum:Reduction_Toom_Cook_interpolation}
    \item Return $Q(M,1) = P_1(M,1) \cdot P_2(M,1) = a \cdot b$. \label{enum:Reduction_Toom_Cook_return_product}
\end{enumerate}

\subsubsection{Correctness}
\label{sec:reduction-correctness}

\begin{theorem} The Toom--Cook rational-to-Gaussian integer multiplication reduction scheme from \S\ref{sec:reduction-scheme} is well-defined and correct.
\end{theorem}
\begin{proof}
    Since $Q(M,1) = P_1(M,1) \cdot P_2(M,1) = a \cdot b$ by definition, all we need to prove is that $Q$ is reconstructible from the points $((x,y), Q(x,y))$, $(x,y) \in R \sqcup S$. Since $Q \in \mathbb{Z}[X,Y]$, we have $Q(\overline{x},\overline{y}) = \overline{Q(x,y)}$. Hence, we know the value of $Q$ at all the points in $T = R \sqcup S \sqcup \overline{S}$. That $Q$ is reconstructible from these points now follows from the fact that $Q$ has homogeneous degree of at most $2k-2$ and the points in $T$ project onto $2k - 1$ distinct points in $P^1(\mathbb{Q}(i))$.
\end{proof}

\subsubsection{Complexity}
\label{sec:reduction-complexity}

\begin{theorem}
    \label{thm:red-complexity}
    Let $\Mult_{\mathbb{Z}}$ be the integer multiplication algorithm from the Toom--Cook rational-to-Gaussian integer multiplication reduction scheme in \S\ref{sec:reduction-scheme} based on a Gaussian integer multiplication algorithm $\Mult_{\mathbb{Z}[i]}$. Suppose the arithmetic complexity of $\Mult_{\mathbb{Z}[i]}$ is bounded by $\alpha n^\omega+\beta n$, where $\omega > 1$  and $\alpha, \beta \ge 0$, for two $n$-digit Gaussian integers. Furthermore, assume the additive complexity of the matrix $U$ associated to the evaluation phase \ref{enum:Reduction_Toom_Cook_evaluation} is given by $q_U$ and the additive complexity of the matrix $W^T$ associated to the interpolation phase \ref{enum:Reduction_Toom_Cook_interpolation} is given by $q_{W^T}$. Then, the arithmetic complexity of $\Mult_{\mathbb{Z}}$ is at most
    \begin{equation}
        \begin{cases}
            \frac{\alpha |S|}{k^{\omega} - |R|}n^{\omega} + \frac{\beta|S| + 2q_U + 2q_{W^T} + 2k - 1}{k - |R|}n + n^{\log_k(|R|)}, & |R| < k, \\
            \frac{\alpha |S|}{k^{\omega} - |R|}n^{\omega} + \frac{\beta|S| + 2q_U + 2q_{W^T} + 2k - 1}{k} \log_k(n) n + n^{\log_k(|R|)}, & |R| = k, \\
            \frac{\alpha |S|}{k^{\omega} - |R|}n^{\omega} + \frac{\beta|S| + 2q_U + 2q_{W^T} + |R| + k - 1}{|R| - k}n^{\log_k(|R|)}, & k < |R| < k^{\omega}, \\
            \frac{\alpha |S|}{k^{\omega}} \log_k(n) n^{\omega} + \frac{\beta|S| + 2q_U + 2q_{W^T} + |R| + k - 1}{|R| - k}n^{\log_k(|R|)}, & |R| = k^{\omega}, \\
            \left(\frac{\alpha |S|}{|R| - k^{\omega}} + \frac{\beta|S| + 2q_U + 2q_{W^T} + |R| + k - 1}{|R| - k} \right) n^{\log_k(|R|)}, & k^{\omega} < |R|, \\
        \end{cases}
    \end{equation}
    where the inputs are two $n$-digit rational integers and $n$ is a power of $k$.
\end{theorem}

\begin{proof}
    Let $F_k(n)$ denote the arithmetic complexity of $\Mult_{\mathbb{Z}}$ for two $n$-digit rational integers as inputs. During the recursive phases \ref{enum:Reduction_Toom_Cook_recursive_gauss_multiplication} and \ref{enum:Reduction_Toom_Cook_recursive_multiplication}, the algorithm $\Mult_{\mathbb{Z}}$ invokes $|R|$ calls to $\Mult_{\mathbb{Z}}$ with two $\tfrac{n}{k}$-digit rational integers and $|S|$ calls to $\Mult_{\mathbb{Z}[i]}$ with two $\tfrac{n}{k}$-digit Gaussian integers. Furthermore, the evaluation phase \ref{enum:Reduction_Toom_Cook_evaluation} costs $2q_U\tfrac{n}{k}$ as we are dealing with $\tfrac{n}{k}$-digit rational integers and the interpolation phase \ref{enum:Reduction_Toom_Cook_interpolation} costs $q_{W^T}\tfrac{2n}{k}$. At last, to compute the result \ref{enum:Reduction_Toom_Cook_return_product}, we can split the coefficients into even and odd powers and evaluate the resulting polynomials separately. This is free as it is a concatenation. The final addition is now among two $(k\frac{2n}{k} - \frac{n}{k})$-digit rational integers as we can save the $\frac{n}{k}$ least significant digits. This yields a cost of $(2k-1)\frac{n}{k}$. We arrive at the following recursive estimate for the arithmetic complexity:
    \begin{equation}
        F_k(n) \le |R| F_k\left(\frac{n}{k}\right) + |S| \left( \alpha \left( \frac{n}{k} \right)^{\omega} + \beta \frac{n}{k} \right) + \left(2q_U + 2q_{W^T}+ 2k - 1\right)\frac{n}{k}.
    \end{equation}
    Iterating this inequality, we find
    \begin{equation}
    \begin{aligned}\label{eq:arithm_cost_reduction_factor}
        F_k(n) &\le \sum_{\ell = 0}^{\log_k(n) - 1} \left[ \left( \frac{|R|}{k^{\omega}} \right)^{\ell} \frac{\alpha |S|}{k^{\omega}} n^{\omega}  + \left( \frac{|R|}{k} \right)^{\ell} \frac{\beta|S| + 2q_U + 2q_{W^T} + 2k - 1}{k}n \right] + |R|^{\log_k(n)} F_k(1) \\
        & \le \frac{\left( \frac{|R|}{k^{\omega}} \right)^{\log_k(n)} - 1}{\left( \frac{|R|}{k^{\omega}} \right) - 1} \frac{\alpha |S|}{k^{\omega}} n^{\omega} + \frac{\left( \frac{|R|}{k} \right)^{\log_k(n)} - 1}{\left( \frac{|R|}{k} \right) - 1} \frac{\beta|S| + 2q_U + 2q_{W^T} + 2k - 1}{k}n + n^{\log_k(|R|)} \\
        & \le
        \begin{cases}
        \frac{\alpha |S|}{k^{\omega} - |R|}n^{\omega}, & |R| < k^{\omega}, \\
        \frac{\alpha |S|}{k^{\omega}} \log_k(n) n^{\omega}, & |R| = k^{\omega}, \\
        \frac{\alpha |S|}{|R| - k^{\omega}} n^{\log_k(|R|)}, & |R| > k^{\omega},
        \end{cases}
        + 
        \begin{cases}
        \frac{\beta|S| + 2q_U + 2q_{W^T} + 2k - 1}{k - |R|}n, & |R| < k, \\
        \frac{\beta|S| + 2q_U + 2q_{W^T} + 2k - 1}{k} \log_k(n) n, & |R| = k, \\
        \frac{\beta|S| + 2q_U + 2q_{W^T} + 2k - 1}{|R| - k} n^{\log_k(|R|)}, & |R| > k,
        \end{cases}
        + n^{\log_k(|R|)}.
    \end{aligned}
    \end{equation}
\end{proof}

\begin{corollary}
    \label{cor:reduction}
    With the setting as in Theorem~\ref{thm:red-complexity}, assume $\omega = \log_k(2k-1)$ as it would be in a Gaussian Toom--Cook-$k$ multiplication algorithm and $|R| < 2k - 1$. Then, the arithmetic complexity of $\Mult_{\mathbb{Z}}$ is at most
    \begin{equation}
        \left(\frac{\alpha}{2}+o(1)\right)n^{\log_k(2k - 1)}.
    \end{equation}
\end{corollary}
\begin{proof}
    In the case where $|R|<2k-1=k^\omega$ (case number 1, 2, and 3 of Theorem~\ref{thm:red-complexity}), the dominant term is
    \begin{equation}
        \frac{\alpha |S|}{k^{\log_k(2k - 1)} - |R|}n^{\log_k(2k - 1)},
    \end{equation}
    which after using $|R|+2|S|=2k - 1$ and simplifying yields the result.
\end{proof}

Corollary~\ref{cor:reduction} shows that if we combine a Toom--Cook-$k$ rational-to-Gaussian reduction scheme with a Gaussian Toom--Cook-$k$ integer multiplication algorithm, we fully recuperate the doubled cost of Gaussian addition.

\subsection{Gaussian Toom--Cook}
\label{sec:Gaussian-integer-multiplication}

\subsubsection{Algorithm scheme}
\label{sec:Gaus_alg_decr}
Let $k\in \mathbb{Z}_{>0}$ and $S\subset \mathbb{Z}[i]^2\backslash \{(0,0)\}$ be a set of $2k-1$ pairs of Gaussian integers that project onto $2k-1$ distinct points on the projective line $P^{1}(\mathbb{Q}(i))$, that is, they are not scalar multiples of each other over $\mathbb{Q}(i)^{\times}$. Let $a$ and $b$ be two Gaussian integers. To compute the product $a \cdot b$, we do the following:
\begin{enumerate}[label=(\Roman*)]
    \item Split $a$ and $b$ into $k$ parts each, denoted $[a]_j$, $[b]_j$, $j=0,\dots, k-1$, respectively, such that there is $M\in \mathbb{Z}_{> 0}$ with
    \begin{align}
        a=\sum_{j=0}^{k-1}[a]_j M^j, && b=\sum_{j=0}^{k-1}[b]_j M^j.
    \end{align}
    \item Define the polynomials $P_1, P_2 \in \mathbb{Z}[i][X,Y]$ as
    \begin{align}
            P_1(X,Y)=\sum_{j=0}^{k-1}[a]_jX^j Y^{k-1-j}, &&
            P_2(X,Y)=\sum_{j=0}^{k-1}[b]_jX^j Y^{k-1-j},
    \end{align}
    such that $P_1(M,1)=a$ and $P_2(M,1)=b$.
    \item For every $(x,y)\in S$, compute the pair $(P_1(x,y),P_2(x,y))$. \label{enum:Gaussian_Toom_Cook_evaluation}
    \item For every $(x,y)\in S$, compute the product $Q(x,y)=P_1(x,y) \cdot P_2(x,y)$. If the size of either $P_1(x,y)$ or $P_2(x,y)$ is too big, compute the product recursively using the same procedure. \label{enum:Gaussian_Toom_Cook_recursive_multiplication}
    \item Use interpolation to retrieve the coefficients of $Q = P_1 \cdot P_2$ from $Q(x,y)$, $(x,y)\in S$. \label{enum:Gaussian_Toom_Cook_interpolation}
    \item Return $Q(M,1) = P_1(M,1) \cdot P_2(M,1) = a \cdot b$. \label{enum:Gaussian_Toom_Cook_return_product}
\end{enumerate}

\begin{remark}
    Although we stated the algorithm over the Gaussian integers, it works more generally over an integral domain of characteristic $0$.
\end{remark}



\subsubsection{Correctness}
\label{sec:Gaussian-correctness}

\begin{theorem} The Gaussian Toom--Cook integer multiplication scheme from \S\ref{sec:Gaus_alg_decr} is well-defined and correct.
\end{theorem}
\begin{proof}
    Since $Q(M,1) = P_1(M,1) \cdot P_2(M,1) = a \cdot b$ by definition, all we need to prove is that $Q$ is reconstructible from the points $((x,y), Q(x,y))$, $(x,y) \in S$. That $Q$ is reconstructible from these points follows from the fact that $Q$ has homogeneous degree of at most $2k-2$ and the points in $S$ project onto $2k - 1$ distinct points in $P^1(\mathbb{Q}(i))$.
\end{proof}

\subsubsection{Complexity} \label{sec:complexity-analysis}

The arithmetic complexity of Gaussian Toom--Cook-$k$ is the sum of the arithmetic complexity of the evaluation phase, pointwise multiplication phase, the interpolation, and carry phases. Let $\langle U,V,W\rangle$ be the bilinear form of Gaussian Toom--Cook-$k$. Let $q_U$, $q_V$, and $q_{W^T}$ be the additive complexity of these matrices. 
The following theorem defines the arithmetic complexity of Gaussian Toom--Cook-$k$ algorithm.

\begin{theorem}\label{thm:Gaus_arith_cost}
    Let $\langle U,V,W\rangle$ be the bilinear form of Gaussian Toom--Cook-$k$. Then, the arithmetic complexity of the algorithm is bounded by
    \begin{equation}
        F_k(n) \le \frac{q_U+q_V+2q_{W^T}+10k-8}{k-1}n^{\log_k(2k-1)}, 
    \end{equation}
     where the inputs are two $n$-digit Gaussian integers and $n$ is a power of $k$.
\end{theorem}

\begin{proof}
The proof is a direct generalisation of \cite[\S4.1]{spiizer2022toom}. Following the Toom--Cook algorithm, see \S\ref{sec:Gaus_alg_decr}, the evaluation phase \ref{enum:Gaussian_Toom_Cook_evaluation} costs $(q_U+q_V)\frac{n}{k}$ as it deals with $\frac{n}{k}$-digit Gaussian integers. The multiplication \ref{enum:Gaussian_Toom_Cook_recursive_multiplication} costs $(2k-1)F_k(\frac{n}{k})$ as there are $2k-1$ multiplications of $\frac{n}{k}$-digit Gaussian integers. The interpolation phase \ref{enum:Gaussian_Toom_Cook_interpolation} costs $q_{W^T}\frac{2n}{k}$ as the operations are on $\frac{2n}{k}$-digit Gaussian integers. Finally, to compute the result \ref{enum:Gaussian_Toom_Cook_return_product}, we may split the coefficients into even and odd powers and evaluate the resulting polynomials separately. This is free as it is a concatenation. The final addition is now among two $(k\frac{2n}{k} - \frac{n}{k})$-digit Gaussian integers as we can save the $\frac{n}{k}$ least significant digits. This yields a cost of $2(2k-1)\frac{n}{k}$.

The resulting recursive relationship is thus:
\begin{equation}
F_k(n) \le  (2k - 1) F_k\left( \frac{n}{k} \right)  + (q_U + q_V) \frac{n}{k} + q_{W^T} \frac{2n}{k} + 2 (2k-1) \frac{n}{k}.
\end{equation}
Iterating, we find
\begin{equation} \begin{aligned}
F_k(n) &\le \sum_{\ell = 0}^{\log_k(n)-1} \left(\frac{2k-1}{k}\right)^{\ell} \frac{q_U + q_V + 2q_{W^T} + 4k - 2}{k}  n + (2k-1)^{\log_k(n)}F_k(1)\\
&\le \frac{ \left(\frac{2k-1}{k}\right)^{\log_k(n)} - 1 }{\left(\frac{2k-1}{k}\right) - 1 } \cdot \frac{q_U + q_V + 2q_{W^T} + 4k - 2}{k} n  + 6n^{\log_k(2k-1)}\\
&\le \frac{q_U + q_V + 2q_{W^T} + 10k - 8}{k-1} n^{\log_{k}(2k-1)} .
\end{aligned}
\end{equation}
We conclude the proof.
\end{proof}

	\section{Steiner trees}
\label{sec:Steiner-tree}

Bodrato and Zanoni~\cite{bodrato} introduced the Toom-graph technique in order to find efficient interpolation schemes for Toom--Cook algorithms.
They formulated the problem as a weighted shortest path problem on the graph whose vertices are given by invertible matrices and the edges are given by elementary row operations.

Here, we generalise this approach to a Steiner-tree problem~\cite{steiner-tree-problems} which not only applies to the interpolation sequence, but also the fast evaluation problem. Steiner trees represent a class of optimisation problems focused on constructing a minimum-weight network that connects a designated set of terminal nodes \cite{steiner-tree-problems}. Although historically employed for network topology and VLSI routing in Euclidean and rectilinear spaces \cite{steiner-tree-problems}, extending this paradigm to directed hypergraphs establishes a rigorous framework for modelling computational sequences \cite{gallo-hypergraphs}, thus suitable for our goal.

\subsection{Problem statement}

The base object is a weighted directed hypergraph $H=(V,E,\omega)$ whose vertices are vectors in a field $K$, that is $V=K^n$, for some $n \in \mathbb{Z}_{>0}$. For us, the underlying field is $K=\mathbb{Q}$ or $\mathbb{Q}(i)$. The hyperedges $e$ are either one-to-one or two-to-one and correspond to vector operations:
\begin{align}
    e&=(\{u\}, \{w\}) \in E & &\Longleftrightarrow & & \exists \lambda \in K: w = \lambda u, \\
    e&=(\{u, v\}, \{w\}) \in E & &\Longleftrightarrow && w = \pm u \pm v.
\end{align}
The weight of an edge is the corresponding cost in the cost model from \S\ref{sec:computational-model}. We note that this makes $H$ a weighted (binary) B-hypergraph~\cite{gallo-hypergraphs}. In order to formalise the general problem we need a notion of reachability.

\begin{definition}
    A vertex $v \in V$ is said to be \emph{reachable} from a set of vertices $U \subseteq V$ through a set of hyperedges $F \subseteq E$ if and only if there is a sequence of vertices $w_0, w_1, \dots, w_{\ell}$ and hyperedges $f_0, f_1, \dots, f_{\ell}$ such that $w_{\ell} = v$ and $f_i = (U_i, \{w_i\}) \in F$ with $U_i \subseteq U \cup \{ w_0, w_1, \dots, w_{i-1} \}$ for $i = 0, 1, \dots, \ell$.
\end{definition}

The optimisation problem can now be stated as follows.

\begin{problem} \label{prob:minimal-reachability}
    Given a set of input vertices $U \subseteq V$ and output vertices $W \subseteq V$ of a weighted hypergraph $H=(V,E,\omega)$, determine a set of hyperedges $F \subseteq E$ of least combined weight such that every vertex $w \in W$ is reachable from $U$ through $F$.
\end{problem}

This problem was shown to be NP-hard for general B-hypergraphs by Italiano--Nanni~\cite[Thm.\@ 3.1]{italiano1989online}. The restriction to B-hypergraphs with the source set of hyperedges restricted to at most 2 vertices remains NP-hard. The reason is that one can replace a hyperedge $e=(\{v_0,v_1,\dots,v_{\ell}\}, \{w\})$ of weight $\omega(e)$ with a series of hyperedges and dummy vertices $d_{0,1,\dots, i}$:
\begin{equation}
    (\{v_0,v_1\}, \{d_{0,1}\}), (\{d_{0,1},v_2\}, \{d_{0,1,2}\}), \dots, (\{d_{0,1, \dots, \ell -1 },v_{\ell}\}, \{d_{0,1, \dots, \ell }\}), \quad \text{and} \quad (\{d_{0,1, \dots, \ell }\}, \{w\}) 
\end{equation}
all of which have weight $0$, except the last one which has weight $\omega(e)$. It was furthermore shown by Boyar--Matthews--Peralta~\cite{boyar-matthews-peralta-shortest-linear-program} that a subproblem is MaxSNP-complete and, hence, also APX-complete.

\subsection{Comparison to Toom graph}

The biggest difference between the Steiner-tree and the Toom-graph technique is that the Steiner tree is more general. It allows one to reuse already computed intermediate steps whereas the Toom graph does not. %
This extra flexibility is not just for the sake of generality as it is strictly needed in the most efficient Gaussian Toom--Cook-$5$ interpolation sequence we have found.\footnote{The black pebble game~\cite{sethi1973complete} requires at least $10$ pebbles for this interpolation sequence.} %
Furthermore, Steiner trees are not restricted by basis-preserving transformation and can thus also be used to find efficient evaluation sequences on top of finding efficient interpolation sequences. On the other hand, the problem to solve on the Toom graph is a simple shortest path, which can be solved in polynomial time, whereas our Steiner-tree problem is NP-hard.

\subsection{Heuristics}

As is the case for the Toom graph technique, the actual hypergraph on which the problem is stated is infinite and thus infeasible to solve completely directly. Thus, like Bodrato and Zanoni~\cite{bodrato}, we introduce heuristics which generate a sub-hypergraph on which we can solve the problem.

Our general heuristic framework is based on ideas from beam-search~\cite{Lowerre1976Harpy}, $A^{\ast}$~\cite{Poh70}, and Boyar--Peralta heuristics~\cite{BMP13}. We keep a beam of partial solutions which are scored based on the number of operations already taken and an estimate for the remaining operations. We implemented two variants for the remainder: a lower bound and an upper bound.

\begin{itemize}
    \item For the lower bound, we use the machinery of Bodrato--Zanoni~\cite[Thm.\@ 3]{bodrato} to get a bound on the number of additions. However, for the number of divisions, we divert from their bound and use the structure given by the elementary divisor theorem. In \S\ref{sec:division-gaussian-toom-cook}, we shall prove that every elementary divisor which is not a unit gives rise to a division in the interpolation phase. A cheap way to (under)estimate the non-unit ones is to count the ones which are not divisible by $(1+i)$. This we achieve by reducing the integral span of the vectors modulo $(1+i)$ and compute the dimension of their null-space over $\mathbb{F}_2$. The latter is efficiently computed using Gaussian elimination using only XOR operations.
    \item For the upper bound, we use a guided Gaussian elimination strategy targeting eliminations which are either cheap in arithmetic cost, generate the most new zero entries, or make a row divisible by a common factor, thus preparing for a division.
    \item To escape local minima, we run the search algorithm iteratively, where between iterations we try to find small optimizations to the solution using AI. This knowledge of a better solution updates the heuristic and allows escaping solutions that are local minima from the perspective of the heuristic.
\end{itemize}

To improve the overall search quality and speed, we also make use of symmetry. Since multiplying by a unit is free, we can reduce each vector modulo units. Furthermore, we include in the initial beam the butterfly moves found in a fast Fourier transform as they are an efficient set of operations.

\subsection{ILP formulation}

In order to solve the minimisation problem on the subhypergraph provided by the heuristics, we formulate the problem as an integer linear programme (ILP). Let $H=(V,E, \omega)$ be the hypergraph produced by the heuristics. Further, let $I$ be the inputs and $O$ the outputs. For the ILP, we need the following variables:
\begin{align}
    x_e &\in \{0,1\}, &&  \forall e \in E, & &\text{representing membership in the solution }F,\\
    y_v &\in \{0,1\}, &&  \forall v \in V, &&\text{representing reachability},\\
    t_v &\in \mathbb{R}_{\ge 0}, &&  \forall v \in V, &&\text{used for acyclic constraints}.
\end{align}
We also need the following (in)equalities:
\begin{align}
    y_v &= 1, &&\forall v \in I \cup O, \label{eq:ilp-input-output-reach}\\
    y_w &\le \sum_{(U,\{w\}) \in E} x_{(U,\{w\})}, && \forall w \in V, w \not \in I, \label{eq:ilp-or}\\
    x_e &\le y_u, && \forall e=(U, \{w\}) \in E, \forall u \in U, \label{eq:ilp-and}\\
    t_v &= 0, && \forall v \in I,\\
    t_u + 1 - M(1-x_e) &\le t_w, && \forall e=(U,\{w\})\in E, \forall u \in U \label{eq:ilp-acyclic},
\end{align}
for some sufficiently large constant $M$. Condition~\eqref{eq:ilp-input-output-reach} ensures that the input vertices are reachable and the output vertices must be reachable. Inequality~\eqref{eq:ilp-or} ensures that one of the incoming hyperedges of a reachable vertex must be in the solution set. Inequality~\eqref{eq:ilp-and} ensures that a hyperedge is only usable for reachability if all of its input vertices are reachable. Note that the weights are non-negative and thus there is a minimal solution which avoids these unnecessary hyperedges. The last inequality, Inequality~\eqref{eq:ilp-acyclic}, ensures that reachability originate from the input vertices and that there is no cyclic reachability argument. That is arguments such as, $v$ is reachable because $w$ is reachable and vice versa. Finally, the ILP optimisation problem is to minimise:
\begin{equation}
    \sum_{e \in E} \omega(e) x_e.
\end{equation}
	\section{Algorithm details}
\label{sec:alg-details}

In this section, we provide detailed descriptions of the evaluation points, the evaluation schemes, and interpolation schemes for both rational-to-Gaussian Toom--Cook-$k$ and Gaussian Toom--Cook-$k$ for $k=3, 4, 5$.

In order to find these solutions, we used a combination of hand optimisations, AI assisted search, Steiner trees from \S\ref{sec:Steiner-tree}, and the heuristic framework around it. All of which were instrumental to getting the arithmetic cost as low as possible.


Throughout this section, we will make use of row operations on matrices for which we shall introduce some notation. We represent the $i$-th (starting from $0$) row by $R_i$ and indicate row operations by `$\gets$'. For example, $R_1 \gets R_1 + iR_2$ indicates, that the second row is replaced by the sum of the second row and the third row multiplied by the scalar $i$. When a step contains multiple operations at once, they should be applied all at once with the right-hand side of `$\gets$' denoting the values of the rows before the current step.

\subsection{Toom--Cook reduction, \texorpdfstring{$k=3$}{k=3}}

\subsubsection{Evaluation points}

An efficient set of points for the reduction is given by
\begin{align}
    R &= \{(0,1), (1,1), (1, 0)\}, & S&=\{(i,1)\}.
\end{align}

\subsubsection{Interpolation sequence}

In order to make the reduction scheme more efficient, we make an additional assumption on the Gaussian multiplication algorithm, namely the input and output Gaussian integers are represented as two integers representing the real and imaginary part, respectively. This will make complex conjugate a null operation and likewise taking imaginary and real part. We shall exploit this by directly using the equations for the imaginary and real part of $Q(i,1)$. That is, we use the equations
\begin{align}
    \Re(Q(i,1)) &= \Re(P_1(i,1)\cdot P_2(i,1)), & &\text{and} & \Im(Q(i,1)) &= \Im(P_1(i,1)\cdot P_2(i,1)),
\end{align}
rather than
\begin{align}
    Q(i,1) &= P_1(i,1)\cdot P_2(i,1), & &\text{and} & Q(-i,1) &= \overline{P_1(i,1)\cdot P_2(i,1)}.
\end{align}
This turns the set of equations to solve for the interpolation into a rational system of equations instead of Gaussian. We now present an efficient sequence of operations to solve the system of equations.

{\allowdisplaybreaks
    \begin{align}    \label{eq:interpolation_sequence-3}
    {W^{-T}} = &
    \begin{bmatrix}
        1 &  &  &  & \\
        1 & 1 & 1 & 1 & 1\\
        1 &   & -1 &  & 1\\
          & 1 &  & -1 & \\
          &  &  &  & 1
    \end{bmatrix}
    & \xRightarrow{\substack{R_1 \gets (R_1 - R_2 - R_3) / 2 \\ R_2 \gets R_0 + R_4 - R_2}} &
    \begin{bmatrix}
        1 &  &  &  & \\
         &  & 1 & 1 & \\
         &  & 1 &  & \\
         & 1 &  & -1 & \\
         &  &  &  & 1
    \end{bmatrix} \\
    \xRightarrow{R_1 \gets R_1 -R_2} &
    \begin{bmatrix}
        1 &  &  &  & \\
         &  &  & 1 & \\
         &  & 1 &  & \\
         & 1 &  & -1 & \\
         &  &  &  & 1
    \end{bmatrix}
    & \xRightarrow{R_3 \gets R_3 + R_1} &
    \begin{bmatrix}
        1 &  &  &  & \\
         &  &  & 1 & \\
         &  & 1 &  & \\
         & 1 &  &  & \\
         &  &  &  & 1
    \end{bmatrix}.
\end{align}
}%
The interpolation sequence~\eqref{eq:interpolation_sequence-3} contains 6 rational additions/subtractions and one division by 2. Hence, we have $q_{W^T} \le 7$ for the arithmetic complexity of $W^T$. The division factor of $W^T$ is $1$.

\subsubsection{Evaluation sequence}
For the evaluation phase, a rational integer $a\in \mathbb{Z}$ is split into the parts $[a]_0$, $[a]_1$, $[a]_2$ and we need to evaluate
\begin{equation}
    P(X,Y) = [a]_0 Y^2 + [a]_1 XY + [a]_2 X^2, \quad \text{ at } (X,Y) \in \{ (0,1), (1,1), (1,0), (i,1)\}.
\end{equation}
We have
\begin{equation}
\begin{aligned}
    P(0,1) &= [a]_0, & \hspace{5em} && P(1,1) &= [a]_0 + [a]_1 + [a]_2, \\
    P(1,0) &= [a]_2, & \hspace{5em} && \Re(P(i,1)) &= [a]_0 - [a]_2, \\
    \Im(P(i,1)) &= [a]_1.
\end{aligned}
\end{equation}
This involves 3 rational additions/subtractions. Hence, the arithmetic complexity of the evaluation satisfies $q_U \le 3$.

\subsection{Toom--Cook reduction, \texorpdfstring{$k=4$}{k=4}}

\subsubsection{Evaluation points}
We choose the set of points
\begin{align}
    R &= \{(0,1), (1,1), (1, 0)\}, & S&=\{(i,1), (1+i,1)\}.
\end{align}

\subsubsection{Interpolation sequence}
We employ the same optimisation strategy as in the case $k=3$. The interpolation phase looks as follows:
{\allowdisplaybreaks
\begin{align} \label{eq:interpolation_sequence-4}
    W^{-T} = &
    \begin{bmatrix}
        1 &  &  &  & & & \\
        1 & 1 & 1 & 1 & 1 & 1 & 1\\
        1 &   & -1 &  & 1 & & -1\\
          & 1 &  & -1 & & 1 &\\
        1 & 1 & & -2 & -4 & -4 &\\
        & 1 & 2 & 2 & & -4 & -8\\
          &  &  & & & & 1
    \end{bmatrix}
    & \xRightarrow{\substack{R_1 \gets R_1 - R_3 \\ R_4 \gets 5R_0 - R_4 \\ R_5 \gets R_5 + 10R_6}} &
    \begin{bmatrix}
        1 &  &  &  & & & \\
        1 &  & 1 & 2 & 1 &  & 1\\
        1 &   & -1 &  & 1 & & -1\\
          & 1 &  & -1 & & 1 &\\
        4 & -1 & & 2 & 4 & 4 &\\
        & 1 & 2 & 2 & & -4 & 2\\
          &  &  & & & & 1
    \end{bmatrix} \\
    \xRightarrow{R_4 \gets (R_4 + R_5) / 2} &
    \begin{bmatrix}
        1 &  &  &  & & & \\
        1 &  & 1 & 2 & 1 &  & 1\\
        1 &   & -1 &  & 1 & & -1\\
          & 1 &  & -1 & & 1 &\\
        2 &  & 1 & 2 & 2 &  & 1\\
        & 1 & 2 & 2 & & -4 & 2\\
          &  &  & & & & 1
    \end{bmatrix}
    & \xRightarrow{\substack{R_1 \gets R_1 - R_2 \\ R_4 \gets R_4 - R_1}} &
    \begin{bmatrix}
        1 &  &  &  & & & \\
         &  & 2 & 2 &  &  & 2\\
        1 &   & -1 &  & 1 & & -1\\
          & 1 &  & -1 & & 1 &\\
        1 &  &  &  & 1 &  & \\
        & 1 & 2 & 2 & & -4 & 2\\
          &  &  & & & & 1
    \end{bmatrix} \\
    \xRightarrow{\substack{R_1 \gets R_1 / 2 \\ R_2 \gets R_4 - R_2 \\ R_5 \gets R_5 - R_1}} &
    \begin{bmatrix}
        1 &  &  &  & & & \\
         &  & 1 & 1 &  &  & 1\\
         &   & 1 &  &  & & 1\\
          & 1 &  & -1 & & 1 &\\
        1 &  &  &  & 1 &  & \\
        & 1 &  &  & & -4 & \\
          &  &  & & & & 1
    \end{bmatrix}
    & \xRightarrow{\substack{R_1 \gets R_1 - R_2 \\ R_2 \gets R_2 - R_6 \\ R_4 \gets R_4 - R_0}} &
    \begin{bmatrix}
        1 &  &  &  & & & \\
         &  &  & 1 &  &  & \\
         &   & 1 &  &  & & \\
          & 1 &  & -1 & & 1 &\\
         &  &  &  & 1 &  & \\
        & 1 &  &  & & -4 & \\
          &  &  & & & & 1
    \end{bmatrix} \\
    \xRightarrow{\substack{R_3 \gets R_3 + R_1}} &
    \begin{bmatrix}
        1 &  &  &  & & & \\
         &  &  & 1 &  &  & \\
         &   & 1 &  &  & & \\
          & 1 &  &  & & 1 &\\
         &  &  &  & 1 &  & \\
        & 1 &  &  & & -4 & \\
          &  &  & & & & 1
    \end{bmatrix}
    & \!\!\!\! \xRightarrow{\substack{R_5 \gets (R_3 - R_5)/5}} &
    \begin{bmatrix}
        1 &  &  &  & & & \\
         &  &  & 1 &  &  & \\
         &   & 1 &  &  & & \\
          & 1 &  &  & & 1 &\\
         &  &  &  & 1 &  & \\
        &  &  &  & & 1 & \\
          &  &  & & & & 1
    \end{bmatrix} \\
    \xRightarrow{\substack{R_3 \gets R_3 - R_5}} &
    \begin{bmatrix}
        1 &  &  &  & & & \\
         &  &  & 1 &  &  & \\
         &   & 1 &  &  & & \\
          & 1 &  &  & &  &\\
         &  &  &  & 1 &  & \\
        &  &  &  & & 1 & \\
          &  &  & & & & 1
    \end{bmatrix}.
\end{align}
}%
This interpolation sequence contains 14 additions/subtractions, 2 multiplications, 2 divisions by a power of 2, and one other division, all over the rational integers. Thus, the arithmetic complexity of $W^T$ satisfies $q_{W^T} \le 19$. The division factor of $W^T$ is $5$.

\subsubsection{Evaluation sequence}

For the evaluation phase, a rational integer $a\in \mathbb{Z}$ is split into the parts $[a]_0$, $[a]_1$, $[a]_2$, $[a]_3$ and we need to evaluate
\begin{multline}
    P(X,Y) = [a]_0 Y^3 + [a]_1 XY^2 + [a]_2 X^2Y + [a]_3X^3, \\  \text{ at } (X,Y) \in \{ (0,1), (1,1), (1,0), (i,1), (1+i,1)\}.
\end{multline}
%
We have
\begin{equation}
\begin{aligned}
    P(0,1) &= [a]_0, & \hspace{1em} && P(1,0) &= [a]_3, \\
    \Re(P(i,1)) &= [a]_0 - [a]_2, & \hspace{1em} && \Im(P(i,1)) &= [a]_1 - [a]_3, \\
    t_0 &= [a]_2 + [a]_3, & \hspace{1em} && t_1 &= [a]_1 + t_0, \\
    \Re(P(1+i,1)) &= \Im(P(i,1)) + [a]_0 - [a]_3, &  && \Im(P(1+i,1)) &= t_0 + t_1, \\
    P(1,1) &= [a]_0 + t_1.
\end{aligned}
\end{equation}
This involves 8 rational additions/subtractions. Hence, the arithmetic complexity of the evaluation satisfies $q_U \le 8$.

\subsection{Toom--Cook reduction, \texorpdfstring{$k=5$}{k=5}}

\subsubsection{Evaluation points}
We choose the set of points
\begin{align}
    R &= \{(0,1), (1,1), (1, 0)\}, & S&=\{(i,1), (1+i,1), (-1+i,1)\}.
\end{align}

\subsubsection{Interpolation sequence}
We employ the same optimisation strategy as in the cases $k=3, 4$. We treat the interpolation phase as follows:

{\allowdisplaybreaks
\begin{align} \label{eq:interpolation_sequence-5}
    W^{-T} = & 
    \begin{bmatrix}
        1 &    &    &    &    &    &    &    &    \\
        1 &  1 &  1 &  1 &  1 &  1 &  1 &  1 &  1 \\
        1 &    & -1 &    &  1 &    & -1 &    &  1 \\
          &  1 &    & -1 &    &  1 &    & -1 &    \\
        1 &  1 &    & -2 & -4 & -4 &    &  8 & 16 \\
          &  1 &  2 &  2 &    & -4 & -8 & -8 &    \\
        1 & -1 &    &  2 & -4 &  4 &    & -8 & 16 \\
          &  1 & -2 &  2 &    & -4 &  8 & -8 &    \\
          &    &    &    &    &    &    &    &  1 \\
    \end{bmatrix} \\
    \xRightarrow{\substack{R_1 \gets (R_1 - R_2 - R_3)/2 \\ R_2 \gets R_2 - R_0 - R_8 \\ R_4 \gets (2R_0 + 32R_8 - R_4 - R_6)/8 \\ R_5 \gets (R_5 - R_7)/4 \\ R_6 \gets R_4 - R_6 \\ R_7 \gets R_5 + R_7}} &
    \begin{bmatrix}
        1 &    &    &    &    &    &    &    &    \\
          &    &  1 &  1 &    &    &  1 &  1 &    \\
          &    & -1 &    &  1 &    & -1 &    &    \\
          &  1 &    & -1 &    &  1 &    & -1 &    \\
          &    &    &    &  1 &    &    &    &    \\
          &    &  1 &    &    &    & -4 &    &    \\
          &  2 &    & -4 &    & -8 &    & 16 &    \\
          &  2 &    &  4 &    & -8 &    & -16&    \\
          &    &    &    &    &    &    &    &  1 \\
    \end{bmatrix} \\
    \xRightarrow{\substack{R_2 \gets R_4 - R_2 \\ R_6 \gets (R_6 + R_7)/4 \\ R_7 \gets (R_7 - R_6)/8}} &
    \begin{bmatrix}
        1 &    &    &    &    &    &    &    &    \\
          &    &  1 &  1 &    &    &  1 &  1 &    \\
          &    &  1 &    &    &    &  1 &    &    \\
          &  1 &    & -1 &    &  1 &    & -1 &    \\
          &    &    &    &  1 &    &    &    &    \\
          &    &  1 &    &    &    & -4 &    &    \\
          &  1 &    &    &    & -4 &    &    &    \\
          &    &    &  1 &    &    &    & -4 &    \\
          &    &    &    &    &    &    &    &  1 \\
    \end{bmatrix} \\
    \xRightarrow{\substack{R_1 \gets R_1 - R_2 \\ R_5 \gets (R_2 - R_5)/5}} &
    \begin{bmatrix}
        1 &    &    &    &    &    &    &    &    \\
          &    &    &  1 &    &    &    &  1 &    \\
          &    &  1 &    &    &    &  1 &    &    \\
          &  1 &    & -1 &    &  1 &    & -1 &    \\
          &    &    &    &  1 &    &    &    &    \\
          &    &    &    &    &    &  1 &    &    \\
          &  1 &    &    &    & -4 &    &    &    \\
          &    &    &  1 &    &    &    & -4 &    \\
          &    &    &    &    &    &    &    &  1 \\
    \end{bmatrix} \\
    \xRightarrow{\substack{R_2 \gets R_2 - R_5 \\ R_3 \gets R_3 + R_1 \\ R_7 \gets (R_1 - R_7)/5}} &
    \begin{bmatrix}
        1 &    &    &    &    &    &    &    &    \\
          &    &    &  1 &    &    &    &  1 &    \\
          &    &  1 &    &    &    &    &    &    \\
          &  1 &    &    &    &  1 &    &    &    \\
          &    &    &    &  1 &    &    &    &    \\
          &    &    &    &    &    &  1 &    &    \\
          &  1 &    &    &    & -4 &    &    &    \\
          &    &    &    &    &    &    &  1 &    \\
          &    &    &    &    &    &    &    &  1 \\
    \end{bmatrix} \\
    \xRightarrow{\substack{R_1 \gets R_1 - R_7 \\ R_6 \gets (R_3 - R_6)/5}} &
    \begin{bmatrix}
        1 &    &    &    &    &    &    &    &    \\
          &    &    &  1 &    &    &    &    &    \\
          &    &  1 &    &    &    &    &    &    \\
          &  1 &    &    &    &  1 &    &    &    \\
          &    &    &    &  1 &    &    &    &    \\
          &    &    &    &    &    &  1 &    &    \\
          &    &    &    &    &  1 &    &    &    \\
          &    &    &    &    &    &    &  1 &    \\
          &    &    &    &    &    &    &    &  1 \\
    \end{bmatrix},
\end{align}
}%
followed by the final operation $R_3 \gets R_3 - R_6$. This sequence contains 21 rational additions/subtractions, 7 multiplication/divisions by powers of 2, and 3 other divisions over the rational integers. Hence, the arithmetic complexity of the interpolation satisfies $q_{W^T} \le 31$.  The division factor of $W^T$ is $5$.

\subsubsection{Evaluation sequence}

For the evaluation phase, a rational integer $a\in \mathbb{Z}$ is split into the parts $[a]_0$, $[a]_1$, $[a]_2$, $[a]_3$, $[a]_4$ and we need to evaluate
\begin{multline}
    P(X,Y) = [a]_0 Y^4 + [a]_1 XY^3 + [a]_2 X^2Y^2 + [a]_3X^3Y + [a]_4X^4, \\ \text{at } (X,Y) \in \{ (0,1), (1,1), (1,0), (i,1), (1+i,1), (-1+i,1)\}.
\end{multline}
We have
\begin{equation}
\begin{aligned}
    P(0,1) &= [a]_0, & \hspace{5em} && P(1,0) &= [a]_4, \\
    t_0 &= 4[a]_2, & \hspace{5em} && t_1 &= 4[a]_4, \\
    t_2 &= [a]_0 + [a]_4, & \hspace{5em} && t_3 &= [a]_2 + [a]_3, \\
    t_4 &= [a]_0 - t_1, & \hspace{5em} && t_5 &= [a]_1 + t_3, \\
    \Im(P(i,1)) &= [a]_1 - [a]_3, & \hspace{5em} && t_6 &= \Im(P(i,1)) - [a]_3, \\
    \Re(P(i,1)) &= t_2 - [a]_2, & \hspace{5em} && P(1,1) &= t_2 + t_5, \\
    \Re(P(1+i,1)) &= t_4 + t_6, & \hspace{5em} && \Re(P(-1+i,1)) &= t_4 - t_6, \\
    \Im(P(1+i,1)) &= t_3 + t_5, & \hspace{5em} && \Im(P(-1+i,1)) &= \Im(P(1+i,1)) - t_0.
\end{aligned}
\end{equation}

This involves 12 rational additions/subtractions and 2 rational multiplications by powers of 2. Hence, the arithmetic complexity of the evaluation satisfies $q_U \le 14$.

\subsection{Gaussian Toom--Cook, \texorpdfstring{$k=3$}{k=3}}

\subsubsection{Evaluation points}

We choose the evaluation points
\begin{equation} \label{eq:Gaussian-TC3-Points}
    S = \{ (0,1), (-1,1), (1,1), (i,1), (1,0) \}.
\end{equation}

\subsubsection{Interpolation sequence}


We present now a series of elementary row operations to invert the interpolation matrix $W^{-T}$. We presume the sequence to be optimal in arithmetic cost. Some discussion in what sense it is indeed optimal may be found in \S\ref{sec:division-gaussian-toom-cook} and \S\ref{sec:division-factor}.


{\allowdisplaybreaks
\begin{align} \label{eq:interpolation_sequence}
    \begin{bmatrix}
        1 &  &  &  & \\
        1 & -1 & 1 & -1 & 1\\
        1 & 1 & 1 & 1 & 1\\
        1 & i & -1 & -i & 1\\
         &  &  &  & 1
    \end{bmatrix}
    && \xRightarrow{R_1 \gets R_2 - R_1} &&
    \begin{bmatrix}
        1 &  &  &  & \\
         & 2 &  & 2 & \\
        1 & 1 & 1 & 1 & 1\\
        1 & i & -1 & -i & 1\\
         &  &  &  & 1
    \end{bmatrix}
    && \xRightarrow{R_1 \gets R_1/2} &&
    \begin{bmatrix}
        1 &  &  &  & \\
         & 1 &  & 1 & \\
        1 & 1 & 1 & 1 & 1\\
        1 & i & -1 & -i & 1\\
         &  &  &  & 1
    \end{bmatrix} 
    \end{align}
    \begin{align}    
    \xRightarrow{\substack{R_2 \gets R_2 - R_1 \\ R_3 \gets R_3 - iR_1}} &
    \begin{bmatrix}
        1 &  &  &  & \\
         & 1 &  & 1 & \\
        1 &  & 1 &  & 1\\
        1 &  & -1 & -2i & 1\\
         &  &  &  & 1
    \end{bmatrix}
    & \xRightarrow{R_3 \gets R_3 - R_2} &
    \begin{bmatrix}
        1 &  &  &  & \\
         & 1 &  & 1 & \\
        1 &  & 1 &  & 1\\
         &  & -2 & -2i & \\
         &  &  &  & 1
    \end{bmatrix} \\
    \xRightarrow{\substack{R_3 \gets iR_3/2 \\ R_2 \gets R_2 - R_0}} &
    \begin{bmatrix}
        1 &  &  &  & \\
         & 1 &  & 1 & \\
         &  & 1 &  & 1\\
         &  & -i & 1 & \\
         &  &  &  & 1
    \end{bmatrix}
    & \xRightarrow{R_2 \gets R_2 - R_4} &
    \begin{bmatrix}
        1 &  &  &  & \\
         & 1 &  & 1 & \\
         &  & 1 &  & \\
         &  & -i & 1 & \\
         &  &  &  & 1
    \end{bmatrix} \\
    \xRightarrow{R_3 \gets R_3 + iR_2} &
    \begin{bmatrix}
        1 &  &  &  & \\
         & 1 &  & 1 & \\
         &  & 1 &  & \\
         &  &  & 1 & \\
         &  &  &  & 1
    \end{bmatrix}
    & \xRightarrow{R_1 \gets R_1 - R_3} &
    \begin{bmatrix}
        1 &  &  &  & \\
         & 1 &  &  & \\
         &  & 1 &  & \\
         &  &  & 1 & \\
         &  &  &  & 1
    \end{bmatrix}
\end{align}
}%
The above inversion sequence includes $8$ additions/subtractions, $2$ divisions by $2$, and $3$ multiplications by the unit $i$ all performed over the Gaussian integers. This Gaussian Toom--Cook-$3$ does not use any non-trivial divisions. Hence, the arithmetic complexity of the interpolation satisfies $q_{W^T} \le 20$ and its division factor is $1$.

\subsubsection{Evaluation sequence}

Let $a \in \mathbb{Z}[i]$ be a Gaussian integer and let $[a]_0$, $[a]_1$, and $[a]_2$ be the parts of $a$ as described in \S\ref{sec:Gaus_alg_decr}. In order to compute
\begin{multline}
    P(X,Y) = [a]_0 Y^2 + [a]_1 XY + [a]_2 X^2, \\ \text{ evaluated at } (X,Y) \in \{ (0,1), (-1,1), (1,1), (i,1), (1,0)\},
\end{multline}
efficiently,  we write
\begin{equation}
\begin{aligned}
    t & \! \coloneq [a]_0 + [a]_2, & \hspace{5em} &&
    P(0,1) &= [a]_0, \\
    P(-1,1) &= t - [a]_1, & \hspace{5em} &&
    P(1,1) &= t + [a]_1, \\
    P(i,1) &= [a]_0 + i[a]_1 - [a]_2, & \hspace{5em} &&
    P(1,0) &= [a]_2.
\end{aligned}
\end{equation}
Thus, the evaluation phase consists of $5$ additions/subtractions and one multiplication by the unit $i$ all performed over the Gaussian integers. The arithmetic complexity of the evaluation sequence thus satisfies $q_U \le 10$.

\subsubsection{Complexity}

\begin{prop}
    The arithmetic complexity of the Gaussian Toom--Cook-$3$ algorithm using the points $S$ from Equation~\eqref{eq:Gaussian-TC3-Points} is bounded by
    \begin{equation}
        41 n^{\log_3 5},
    \end{equation}
    for the multiplication of two $n$-digit Gaussian integers, when $n$ is a power of $3$.
\end{prop}

\begin{proof}
    We apply Theorem~\ref{thm:Gaus_arith_cost} to find
    \begin{equation}
        F_3(n) \le \frac{10 + 10 + 2\cdot 20 + 10 \cdot 3 - 8}{3 - 1} n^{\log_3 5} = 41 n^{\log_3 5}.
    \end{equation}
\end{proof}

\subsection{Gaussian Toom--Cook, \texorpdfstring{$k=4$}{k=4}}

\subsubsection{Evaluation points}
We choose the evaluation points
\begin{equation} \label{eq:Gaussian-TC4-Points}
    S = \{ (0,1), (1,1), (-1,1), (i,1), (-i,1), (2,1), (1,0) \}.
\end{equation}

\subsubsection{Interpolation sequence}
We invert the Vandermonde matrix by elementary row operations as follows:

{\allowdisplaybreaks
\small
\begin{align}
    W^{-T} = \begin{bmatrix}
        1 &  &  &  &  &  & \\
        1 & 1 & 1 & 1 & 1 & 1 & 1\\
        1 & -1 & 1 & -1 & 1 & -1 & 1\\
        1 & i & -1 & -i & 1 & i & -1\\
        1 & -i & -1 & i & 1 & -i & -1\\
        1 & 2 & 4 & 8 & 16 & 32 & 64\\
        & & &  &  &  & 1
    \end{bmatrix}
    && \xRightarrow{\substack{R_2 \gets R_1 - R_2 \\ R_4 \gets (R_3 - R_4)/(i) \\ R_5 \gets (R_5 - R_0)/2}} 
    \begin{bmatrix}
        1 &  &  &  &  &  & \\
        1 & 1 & 1 & 1 & 1 & 1 & 1\\
         & 2 &  & 2 &  & 2 & \\
        1 & i & -1 & -i & 1 & i & -1\\
         & 2 &  & -2 &  & 2 & \\
         & 1 & 2 & 4 & 8 & 16 & 32 \\
        & & &  &  &  & 1
    \end{bmatrix} \\
    \xRightarrow{\substack{R_2 \gets R_2 - R_4 \\ R_4 \gets R_4 / 2}} 
    \begin{bmatrix}
        1 &  &  &  &  &  & \\
        1 & 1 & 1 & 1 & 1 & 1 & 1\\
         &  &  & 4 &  &  & \\
        1 & i & -1 & -i & 1 & i & -1\\
         & 1 &  & -1 &  & 1 & \\
         & 1 & 2 & 4 & 8 & 16 & 32 \\
        & & &  &  &  & 1
    \end{bmatrix}
    && \xRightarrow{\substack{R_2 \gets R_2 / 4 \\ R_3 \gets R_3 - iR_4 \\ R_5 \gets R_5 - R_2}}
    \begin{bmatrix}
        1 &  &  &  &  &  & \\
        1 & 1 & 1 & 1 & 1 & 1 & 1\\
         &  &  & 1 &  &  & \\
        1 &  & -1 &  & 1 &  & -1\\
         & 1 &  & -1 &  & 1 & \\
         & 1 & 2 &  & 8 & 16 & 32 \\
        & & &  &  &  & 1
    \end{bmatrix} \\
    \xRightarrow{\substack{R_1 \gets R_1 - R_2 - R_3 \\ R_3 \gets R_3 - R_0 \\ R_4 \gets R_4 + R_2}}
    \begin{bmatrix}
        1 &  &  &  &  &  & \\
         & 1 & 2 &  &  & 1 & 2\\
         &  &  & 1 &  &  & \\
         &  & -1 &  & 1 &  & -1\\
         & 1 &  &  &  & 1 & \\
         & 1 & 2 &  & 8 & 16 & 32 \\
        & & &  &  &  & 1
    \end{bmatrix}
    && \xRightarrow{\substack{R_1 \gets (R_1 - R_4)/2 \\ R_5 \gets R_5 - R_1}} 
    \begin{bmatrix}
        1 &  &  &  &  &  & \\
         &  & 1 &  &  &  & 1\\
         &  &  & 1 &  &  & \\
         &  & -1 &  & 1 &  & -1\\
         & 1 &  &  &  & 1 & \\
         &  &  &  & 8 & 15 & 30 \\
        & & &  &  &  & 1
    \end{bmatrix} \\
    \xRightarrow{\substack{R_1 \gets R_1 - R_6 \\ R_3 \gets R_3 + R_1}} 
    \begin{bmatrix}
        1 &  &  &  &  &  & \\
         &  & 1 &  &  &  & \\
         &  &  & 1 &  &  & \\
         &  &  &  & 1 &  & \\
         & 1 &  &  &  & 1 & \\
         &  &  &  & 8 & 15 & 30 \\
        & & &  &  &  & 1
    \end{bmatrix}
    && \xRightarrow{\substack{R_5 \gets (R_5 - 8R_3)/15}} 
    \begin{bmatrix}
        1 &  &  &  &  &  & \\
         &  & 1 &  &  &  & \\
         &  &  & 1 &  &  & \\
         &  &  &  & 1 &  & \\
         & 1 &  &  &  & 1 & \\
         &  &  &  &  & 1 & 2 \\
        & & &  &  &  & 1
    \end{bmatrix} \\
    \xRightarrow{\substack{R_5 \gets R_5 - 2R_6}} 
    \begin{bmatrix}
        1 &  &  &  &  &  & \\
         &  & 1 &  &  &  & \\
         &  &  & 1 &  &  & \\
         &  &  &  & 1 &  & \\
         & 1 &  &  &  & 1 & \\
         &  &  &  &  & 1 &  \\
        & & &  &  &  & 1
    \end{bmatrix}
    && \xRightarrow{\substack{R_4 \gets R_4 - R_5}} 
    \begin{bmatrix}
        1 &  &  &  &  &  & \\
         &  & 1 &  &  &  & \\
         &  &  & 1 &  &  & \\
         &  &  &  & 1 &  & \\
         & 1 &  &  &  &  & \\
         &  &  &  &  & 1 &  \\
        & & &  &  &  & 1
    \end{bmatrix}. \label{eq:GTC_interpolation_sequence_4}
\end{align}
}%
The above inversion sequence includes $17$ additions/subtractions, $6$ divisions/multiplications by powers of $2$, two multiplication by the unit $-i$, and one rational division by a non-unit and non-power of $2$ all performed over the Gaussian integers. Thus, the arithmetic complexity of the interpolation phase satisfies $q_{W^T} \le 48$ and the division factor is $15$.

\subsubsection{Evaluation sequence}

Let $a \in \mathbb{Z}[i]$ be a Gaussian integer and let $[a]_0$, $[a]_1$, $[a]_2$, and $[a]_3$ be the parts of $a$ as described in \S\ref{sec:Gaus_alg_decr}. In order to compute
\begin{multline}
    P(X,Y) = [a]_0 Y^3 + [a]_1 XY^2 + [a]_2 X^2Y + [a]_3X^3, \\ \text{evaluated at } (X,Y) \in \{ (0,1), (1,1), (-1,1), (i,1), (-i,1), (2,1), (1,0)\},
\end{multline}
efficiently, we write


\begin{equation}
\begin{aligned}
    t_0^+ & \! \coloneq [a]_0 + [a]_2, & \hspace{4em} &&
    t_0^- & \! \coloneq [a]_0 - [a]_2, \\
    t_1^+ & \! \coloneq [a]_1 + [a]_3, & \hspace{4em} &&
    t_1^- & \! \coloneq [a]_1 - [a]_3, \\
    P(0,1) &= [a]_0, & \hspace{5em} &&
    P(1,1) &= t_0^+ + t_1^+, \\
    P(-1,1) &= t_0^+ - t_1^+, & \hspace{4em} &&
    P(i,1) &= t_0^- + it_1^-, \\
    P(-i,1) &= t_0^- - it_1^-, & \hspace{4em} &&
    P(1,0) &= [a]_3, \\
    s_0 & \! \coloneq [a]_2 - t_1^- + P(1,1), & \hspace{4em} &&
    s_1 & \! \coloneq s_0 - P(-1,1), \\
    P(2,1) &= 2s_1 + s_0.
\end{aligned}
\end{equation}
Thus, the evaluation phase consists of $12$ additions/subtractions, one multiplication by a power of $2$, and $2$ multiplications by the unit $i$ all performed over the Gaussian integers. Hence, the arithmetic complexity of the evaluation phase satisfies $q_U \le 26$.



\subsubsection{Complexity}

\begin{prop}
    The arithmetic complexity of the Gaussian Toom--Cook-$4$ algorithm using the points $S$ from Equation~\eqref{eq:Gaussian-TC4-Points} is bounded by
    \begin{equation}
        60 n^{\log_4 7},
    \end{equation}
    for the multiplication of two $n$-digit Gaussian integers, when $n$ is a power of $4$.
\end{prop}

\begin{proof}
    We apply Theorem~\ref{thm:Gaus_arith_cost} to find
    \begin{equation}
        F_4(n) \le \frac{26 + 26 + 2\cdot 48 + 10 \cdot 4 - 8}{4 - 1} n^{\log_4 7} = 60 n^{\log_4 7}.
    \end{equation}
\end{proof}

\subsection{Gaussian Toom--Cook, \texorpdfstring{$k=4$}{k=4}, alternative points}

\subsubsection{Evaluation points}
An alternative set of evaluation points for Gaussian Toom--Cook-$4$ is given by
\begin{equation} \label{eq:Gaussian-TC4A-Points}
    S = \{ (0,1), (1,1), (1+i,1), (-1-i,1), (-1+i,1), (1-i,1), (1,0) \}. 
\end{equation}

\subsubsection{Interpolation sequence}
We invert the Vandermonde matrix by elementary row operations as follows:

{\allowdisplaybreaks
\begin{align} \label{eq:interpolation_sequence_4_alt}
    W^{-T} =
    &\begin{bmatrix}
        1 &  &  &  &  &  & \\
        1 & 1 & 1 & 1 & 1 & 1 & 1\\
        1 & 1+i & 2i & -2+2i & -4 & -4-4i & -8i\\
        1 & -1-i & 2i & 2-2i & -4 & 4+4i & -8i\\
        1 & -1+i & -2i & 2+2i & -4 & 4-4i & 8i\\
        1 & 1-i & -2i & -2-2i & -4 & -4+4i & 8i\\
        & & &  &  &  & 1
    \end{bmatrix} \\
    \xRightarrow{\substack{R_2 \gets R_2 + R_3 \\ R_3 \gets R_3 - R_2 \\ R_4 \gets R_4 + R_5 \\ R_5 \gets (R_5 - R_4)/(i)}} 
    &\begin{bmatrix}
        1 &  &  &  &  &  & \\
        1 & 1 & 1 & 1 & 1 & 1 & 1\\
        2 & & 4i & & -8 & & -16i\\
          & -2-2i & & 4-4i & & 8+8i &\\
        2 & & -4i & & -8 & & 16i\\
          & -2-2i & & -4+4i & & 8+8i &\\
        & & &  &  &  & 1
    \end{bmatrix} \\
    \xRightarrow{\substack{R_2 \gets (R_2 - R_4)/(8i) \\ R_3 \gets (R_3 - R_5)\cdot (1+i)/16 \\ R_4 \gets (R_2 + R_4)/4 \\ R_5 \gets (R_3 + R_5)\cdot(1-i)/8}} 
    &\begin{bmatrix}
        1 &  &  &  &  &  & \\
        1 & 1 & 1 & 1 & 1 & 1 & 1\\
         & & 1 & &  & & -4\\
          &  & & 1 & & &\\
        1 & &  & & -4 & & \\
          & -1 & & & & 4 &\\
        & & &  &  &  & 1
    \end{bmatrix} \\
    \xRightarrow{\substack{R_1 \gets R_1 - R_0 - R_3 - R_6 \\ R_2 \gets R_2 + 4R_6 \\ R_4 \gets (R_0-R_4)/4}}
    &\begin{bmatrix}
        1 &  &  &  &  &  & \\
         & 1 & 1 &  & 1 & 1 & \\
         & & 1 & &  & & \\
          &  & & 1 & & &\\
         & &  & & 1 & & \\
          & -1 & & & & 4 &\\
        & & &  &  &  & 1
    \end{bmatrix} \\
    \xRightarrow{\substack{R_1 \gets R_1 - R_2 - R_4 }} 
    &\begin{bmatrix}
        1 &  &  &  &  &  & \\
         & 1 &  &  &  & 1 & \\
         & & 1 & &  & & \\
          &  & & 1 & & &\\
         & &  & & 1 & & \\
          & -1 & & & & 4 &\\
        & & &  &  &  & 1
    \end{bmatrix}.
\end{align}
}
Followed by the final operations
\begin{align}
R_5 &\gets (R_5 + R_1) / 5, & R_1 &\gets R_1 - R_5. 
\end{align}
The above inversion sequence includes $17$ additions/subtractions, $6$ divisions/multiplications by powers of $2$, $2$ multiplication by the unit $i$, $2$ multiplication by $1\pm i$, and $1$ divisions by non-unit and non-power of $2$ rational integer all performed over the Gaussian integers. This Gaussian Toom--Cook-$4$ uses one non-trivial division, its arithmetic complexity is $q_{W^T} \le 52$, and its division factor is $5$.

\subsubsection{Evaluation sequence}

Let $a \in \mathbb{Z}[i]$ be a Gaussian integer and let $[a]_0$, $[a]_1$, $[a]_2$, and $[a]_3$ be the parts of $a$ as described in \S\ref{sec:Gaus_alg_decr}. In order to compute
\begin{multline}
    P(X,Y) = [a]_0 Y^3 + [a]_1 XY^2 + [a]_2 X^2Y + [a]_3X^3, \\ \text{evaluated at } (X,Y) \in \{ (0,1), (1,1), (1+i,1), (-1-i,1), (-1+i,1), (1-i,1), (1,0)\},
\end{multline}
efficiently, we write

\begin{equation}
\begin{aligned}
    s_2 & \! \coloneq 2i[a]_2, & \hspace{0em} &&
    s_3 & \! \coloneq 2[a]_3, \\
    t_0^+ & \! \coloneq [a]_0 + s_2, & \hspace{0em} &&
    t_0^- & \! \coloneq [a]_0 - s_2, \\
    t_1^+ & \! \coloneq i([a]_1 + s_3), & \hspace{0em} &&
    t_1^- & \! \coloneq [a]_1 - s_3, \\
    t_2^+ & \! \coloneq t_1^+ + t_1^-, & \hspace{0em} &&
    t_2^- & \! \coloneq t_1^+ - t_1^-, \\
    P(0,1) &= [a]_0, & \hspace{0em} &&
    P(1,1) &= [a]_0 + [a]_1 + [a]_2 + [a]_3, \\
    P(1+i,1) &= t_0^+ + t_2^+, & \hspace{0em} &&
    P(-1-i,1) &= t_0^+ - t_2^+, \\
    P(-1+i,1) &= t_0^- + t_2^-, & \hspace{0em} &&
    P(1-i,1) &= t_0^- - t_2^-, \\
    P(1,0) &= [a]_3.
\end{aligned}
\end{equation}
Thus, the evaluation phase consists of $13$ additions/subtractions, $2$ multiplications by a power of $2$, and $2$ multiplications by the unit $i$ all performed over the Gaussian integers. Hence, the arithmetic complexity of the evaluation phase satisfies $q_U \le 30$.

\subsubsection{Complexity}

\begin{prop}
    The arithmetic complexity of the Gaussian Toom--Cook-$4$ algorithm using the points $S$ from Equation~\eqref{eq:Gaussian-TC4A-Points} is bounded by
    \begin{equation}
        \frac{196}{3} n^{\log_4 7},
    \end{equation}
    for the multiplication of two $n$-digit Gaussian integers, when $n$ is a power of $4$.
\end{prop}

\begin{proof}
    We apply Theorem~\ref{thm:Gaus_arith_cost} to find
    \begin{equation}
        F_4(n) \le \frac{30 + 30 + 2\cdot 52 + 10 \cdot 4 - 8}{4 - 1} n^{\log_4 7} = \frac{196}{3} n^{\log_4 7}.
    \end{equation}
\end{proof}

\subsection{Gaussian Toom--Cook, \texorpdfstring{$k=5$}{k=5}}

\subsubsection{Evaluation points}
We choose the evaluation points
\begin{equation} \label{eq:Gaussian-TC5-Points}
    S = \{ (0,1), (1,1), (-1,1), (i,1), (-i,1), (1+i,1), (1-i,1), (-1+i,1), (1,0) \}.
\end{equation}

\subsubsection{Interpolation sequence}
We invert the Vandermonde matrix by elementary row operations as follows:

In addition to the rows $R_0,\dots,R_8$ we use one auxiliary register $R_9$, which is appended as a tenth row in the fourth step.

{ \allowdisplaybreaks
 \begin{align} \label{eq:GTC-interpolation_sequence_5}
    W^{-T} = & \begin{bmatrix}
        1 &  &  &  &  &  &  &  & \\
        1 & 1 & 1 & 1 & 1 & 1 & 1 & 1 & 1\\
        1 & -1 & 1 & -1 & 1 & -1 & 1 & -1 & 1\\
        1 & i & -1 & -i & 1 & i & -1 & -i & 1\\
        1 & -i & -1 & i & 1 & -i & -1 & i & 1\\
        1 & 1+i & 2i & -2+2i & -4 & -4-4i & -8i & 8-8i & 16\\
        1 & 1-i & -2i & -2-2i & -4 & -4+4i & 8i & 8+8i & 16\\
        1 & -1+i & -2i & 2+2i & -4 & 4-4i & 8i & -8-8i & 16\\
         &  &  &  &  &  &  &  & 1
    \end{bmatrix} \\
     \xRightarrow{\substack{R_1 \gets R_1 + R_2 \\ R_2 \gets R_1 - R_2 \\ R_3 \gets R_3 + R_4 \\ R_4 \gets R_3 - R_4}} & \begin{bmatrix}
        1 &  &  &  &  &  &  &  & \\
        2 &  & 2 &  & 2 &  & 2 &  & 2\\
         & 2 &  & 2 &  & 2 &  & 2 & \\
        2 &  & -2 &  & 2 &  & -2 &  & 2\\
         & 2i &  & -2i &  & 2i &  & -2i & \\
        1 & 1+i & 2i & -2+2i & -4 & -4-4i & -8i & 8-8i & 16\\
        1 & 1-i & -2i & -2-2i & -4 & -4+4i & 8i & 8+8i & 16\\
        1 & -1+i & -2i & 2+2i & -4 & 4-4i & 8i & -8-8i & 16\\
         &  &  &  &  &  &  &  & 1
    \end{bmatrix} \\
     \xRightarrow{\substack{R_1 \gets R_1 + R_3 \\ R_2 \gets R_2 - iR_4 \\ R_3 \gets R_1 - R_3 \\ R_4 \gets R_2 + iR_4}} & \begin{bmatrix}
        1 &  &  &  &  &  &  &  & \\
        4 &  &  &  & 4 &  &  &  & 4\\
         & 4 &  &  &  & 4 &  &  & \\
         &  & 4 &  &  &  & 4 &  & \\
         &  &  & 4 &  &  &  & 4 & \\
        1 & 1+i & 2i & -2+2i & -4 & -4-4i & -8i & 8-8i & 16\\
        1 & 1-i & -2i & -2-2i & -4 & -4+4i & 8i & 8+8i & 16\\
        1 & -1+i & -2i & 2+2i & -4 & 4-4i & 8i & -8-8i & 16\\
         &  &  &  &  &  &  &  & 1
    \end{bmatrix} \\
     \xRightarrow{\substack{R_1 \gets R_1/4 - R_0 - R_8}} & \begin{bmatrix}
        1 &  &  &  &  &  &  &  & \\
         &  &  &  & 1 &  &  &  & \\
         & 4 &  &  &  & 4 &  &  & \\
         &  & 4 &  &  &  & 4 &  & \\
         &  &  & 4 &  &  &  & 4 & \\
        1 & 1+i & 2i & -2+2i & -4 & -4-4i & -8i & 8-8i & 16\\
        1 & 1-i & -2i & -2-2i & -4 & -4+4i & 8i & 8+8i & 16\\
        1 & -1+i & -2i & 2+2i & -4 & 4-4i & 8i & -8-8i & 16\\
         &  &  &  &  &  &  &  & 1
    \end{bmatrix} \\
     \xRightarrow{\substack{R_9 \gets 2(R_0 - 4(R_1 - 4R_8))}} & \begin{bmatrix}
        1 &  &  &  &  &  &  &  & \\
         &  &  &  & 1 &  &  &  & \\
         & 4 &  &  &  & 4 &  &  & \\
         &  & 4 &  &  &  & 4 &  & \\
         &  &  & 4 &  &  &  & 4 & \\
        1 & 1+i & 2i & -2+2i & -4 & -4-4i & -8i & 8-8i & 16\\
        1 & 1-i & -2i & -2-2i & -4 & -4+4i & 8i & 8+8i & 16\\
        1 & -1+i & -2i & 2+2i & -4 & 4-4i & 8i & -8-8i & 16\\
         &  &  &  &  &  &  &  & 1\\
        2 &  &  &  & -8 &  &  &  & 32
    \end{bmatrix} \\
     \xRightarrow{\substack{R_5 \gets R_5 - R_9}} & \begin{bmatrix}
        1 &  &  &  &  &  &  &  & \\
         &  &  &  & 1 &  &  &  & \\
         & 4 &  &  &  & 4 &  &  & \\
         &  & 4 &  &  &  & 4 &  & \\
         &  &  & 4 &  &  &  & 4 & \\
        -1 & 1+i & 2i & -2+2i & 4 & -4-4i & -8i & 8-8i & -16\\
        1 & 1-i & -2i & -2-2i & -4 & -4+4i & 8i & 8+8i & 16\\
        1 & -1+i & -2i & 2+2i & -4 & 4-4i & 8i & -8-8i & 16\\
         &  &  &  &  &  &  &  & 1\\
        2 &  &  &  & -8 &  &  &  & 32
    \end{bmatrix} \\
     \xRightarrow{\substack{R_6 \gets R_6 + R_5 \\ R_7 \gets i(R_7 + R_5) \\ R_9 \gets (iR_3 - R_9 + R_6 + R_7)/(20i)}} & \begin{bmatrix}
        1 &  &  &  &  &  &  &  & \\
         &  &  &  & 1 &  &  &  & \\
         & 4 &  &  &  & 4 &  &  & \\
         &  & 4 &  &  &  & 4 &  & \\
         &  &  & 4 &  &  &  & 4 & \\
        -1 & 1+i & 2i & -2+2i & 4 & -4-4i & -8i & 8-8i & -16\\
         & 2 &  & -4 &  & -8 &  & 16 & \\
         & -2 &  & -4 &  & 8 &  & 16 & \\
         &  &  &  &  &  &  &  & 1\\
         &  &  &  &  &  & 1 &  & 
    \end{bmatrix} \\
     \xRightarrow{\substack{R_3 \gets R_3/4 - R_9 \\ R_5 \gets (R_2 - R_6 + R_7)/20 \\ R_6 \gets (2R_4 + R_6 + R_7)/40}} & \begin{bmatrix}
        1 &  &  &  &  &  &  &  & \\
         &  &  &  & 1 &  &  &  & \\
         & 4 &  &  &  & 4 &  &  & \\
         &  & 1 &  &  &  &  &  & \\
         &  &  & 4 &  &  &  & 4 & \\
         &  &  &  &  & 1 &  &  & \\
         &  &  &  &  &  &  & 1 & \\
         & -2 &  & -4 &  & 8 &  & 16 & \\
         &  &  &  &  &  &  &  & 1\\
         &  &  &  &  &  & 1 &  & 
    \end{bmatrix} \\
     \xRightarrow{\substack{R_2 \gets R_2/4 - R_5 \\ R_4 \gets R_4/4 - R_6}} & \begin{bmatrix}
        1 &  &  &  &  &  &  &  & \\
         &  &  &  & 1 &  &  &  & \\
         & 1 &  &  &  &  &  &  & \\
         &  & 1 &  &  &  &  &  & \\
         &  &  & 1 &  &  &  &  & \\
         &  &  &  &  & 1 &  &  & \\
         &  &  &  &  &  &  & 1 & \\
         & -2 &  & -4 &  & 8 &  & 16 & \\
         &  &  &  &  &  &  &  & 1\\
         &  &  &  &  &  & 1 &  & 
    \end{bmatrix}.
\end{align}
} 

The inversion sequence is concluded as we have generated all elementary vectors. The extra vector can just be omitted.
The above inversion sequence includes $25$ additions/subtractions, $8$ divisions/multiplications by powers of $2$, $5$ multiplications by a unit, and $3$ rational division by a non-unit and non-power of $2$ all performed over the Gaussian integers. Thus, the arithmetic complexity of the interpolation phase satisfies $q_{W^T} \le 72$ and the division factor is $5$.

\subsubsection{Evaluation sequence}

Let $a \in \mathbb{Z}[i]$ be a Gaussian integer and let $[a]_0$, $[a]_1$, $[a]_2$, $[a]_3$, and $[a]_4$ be the parts of $a$ as described in \S\ref{sec:Gaus_alg_decr}. In order to compute
\begin{multline}
    P(X,Y) = [a]_0 Y^4 + [a]_1 XY^3 + [a]_2 X^2Y^2 + [a]_3X^3Y + [a]_4X^4, \text{ evaluated at } \\ (X,Y) \in \{ (0,1), (1,1), (-1,1), (i,1), (-i,1), (1+i,1), (1-i,1), (-1+i,1), (1,0)\},
\end{multline}
efficiently, we write


{
\allowdisplaybreaks
\begin{equation}
\begin{aligned}
    P(0, 1) &= [a]_0, & \hspace{2em} &&
    P(1, 0) &= [a]_4, \\
    t_0 &= [a]_1 - [a]_3, & \hspace{2em} &&
    t_1 &= [a]_1 + [a]_3, \\
    t_2 &= [a]_0 + [a]_4, & \hspace{2em} &&
    t_3 &= t_2 - [a]_2, \\
    t_4 &= t_2 + [a]_2, & \hspace{2em} &&
    t_5 &= [a]_0 - 4[a]_4, \\
    P(1, 1) &= t_4 + t_1, & \hspace{2em} &&
    P(-1, 1) &= t_4 - t_1, \\
    P(i, 1) &= t_3 + it_0, & \hspace{2em} &&
    P(-i, 1) &= t_3 - it_0, \\
    t_6 &= P(1,1) - P(-i, 1), & \hspace{2em} &&
    t_7 &= i(P(1,1) - P(i, 1)), \\
    t_8 &= (-1+i)[a]_3, & \hspace{2em} &&
    t_9 &= t_6 - t_8, \\
    P(-1+i, 1) &= i(t_9 - 4[a]_2) + t_5, & \hspace{2em} &&
    P(1+i, 1) &= t_7 + t_8 + t_5, \\
    P(1-i, 1) &= -it_9 + t_5.
\end{aligned}
\end{equation}
} 
Thus, the evaluation phase consists of $19$ additions/subtractions, $2$ multiplications by a power of $2$, and $6$ multiplications by the unit $i$ all performed over the Gaussian integers. Thus, the arithmetic complexity of the evaluation phase satisfies $q_{U} \le 42$.

\subsubsection{Complexity}

\begin{prop}
    The arithmetic complexity of the Gaussian Toom--Cook-$5$ algorithm using the points $S$ from Equation~\eqref{eq:Gaussian-TC5-Points} is bounded by
    \begin{equation}
        \frac{135}{2} n^{\log_5 9},
    \end{equation}
    for the multiplication of two $n$-digit Gaussian integers, when $n$ is a power of $5$.
\end{prop}

\begin{proof}
    We apply Theorem~\ref{thm:Gaus_arith_cost} to find
    \begin{equation}
        F_5(n) \le \frac{42 + 42 + 2\cdot 72 + 10 \cdot 5 - 8}{5 - 1} n^{\log_5 9} = \frac{135}{2} n^{\log_5 9}.
    \end{equation}
\end{proof}
	\section{Divisions in Gaussian Toom--Cook algorithms}
\label{sec:division-gaussian-toom-cook}

Recall the Gaussian Toom--Cook-$k$ algorithm from \S\ref{sec:Gaus_alg_decr}. In this section, we will prove Theorem~\ref{thm:main-divisions} which states that the Gaussian Toom--Cook-$k$ algorithm requires many divisions for large $k$. In particular, we will show that the interpolation phase \ref{enum:Gaussian_Toom_Cook_interpolation} requires many divisions. The proof will bootstrap from a single division for small $k$, which is proven in \S\ref{sec:divisions-lower-bound-simple}. The full proof of Theorem~\ref{thm:main-divisions} is then concluded in \S\ref{sec:divisions-linear-lower-bound}. The presented proofs are extensions of those given in \cite{Toom-Cook-divisions}, which treats the rational case.

In \S\ref{sec:discussion-interpolcation-revisited}, we revisit the interpolation phase \ref{enum:Gaussian_Toom_Cook_interpolation} and introduce some further structure required in the proofs.

Throughout this section, we assume the reader is familiar with basic concepts in commutative algebra and algebraic number theory, see for example \cite{Bosch-Alg, atiyah-macdonald-commutative-algebra}, respectively \cite{Neukirch-Algebraic-NT}, for relevant background.


\subsection{Interpolation phase revisited} \label{sec:discussion-interpolcation-revisited}

Recall the notation from \S\ref{sec:Gaus_alg_decr}. In the interpolation phase \ref{enum:Gaussian_Toom_Cook_interpolation}, we need to reconstruct the coefficients $q_j$, $j=0,\dots,2k-2$, of the polynomial $Q$ from a set of evaluation points $S$ of cardinality $2k-1$. This may be phrased as a system of linear equations:
\begin{equation} \label{eq:general-interpolation-linear-system}
    \begin{bmatrix} Q(x,y) \end{bmatrix}_{(x,y) \in S} = V(S) \begin{bmatrix} q_j \end{bmatrix}_{0 \le j \le 2k-2},
\end{equation}
where
\begin{equation} \label{eq:Vandermonde-abbreviation}
    V(S) = \begin{bmatrix} x^j y^{|S|-1-j} \end{bmatrix}_{(x,y)\in S,\, 0 \le j \le |S|-1}
\end{equation}
is a Vandermonde matrix with coefficients in $\mathbb{Z}[i]$. The Vandermonde matrix \eqref{eq:Vandermonde-abbreviation} is invertible in $\mathbb{Q}(i)$ as the evaluation points project to $2k-1$ distinct points on the projective line $P^1(\mathbb{Q}(i))$ by assumption. This system of linear equations may be solved by Gaussian elimination as was done in \S\ref{sec:alg-details}, that is, applying elementary row operations.

Another way of looking at the process of solving for the coefficients $q_j$, for $j=0,\dots,2k-2$, is as a sequence of vectors in the vector space $W=\bigoplus_{j=0}^{2k-1} \mathbb{Q}(i)q_j$. Here, we abused notation to treat $q_j$, for $j=0,\dots,2k-2$, as the standard basis vectors of the aforementioned vector space. We start with the set $\mathcal{W}_0 \subseteq W$ containing the row vectors of $V(S)$, which correspond to $Q(x,y)$ for $(x,y) \in S$. Each element $w_i$ of the sequence is computed as either
\begin{enumerate}[label=(\roman*)]
    \item $w_i = \lambda v + \mu u$ for $\lambda, \mu \in \mathbb{Z}[i]$ and $v, u \in \mathcal{W}_{i-1}$, or \label{enum:sequence-vector-int-linear-combination}
    \item $w_i = v/\lambda$ for $\lambda \in \mathbb{Z}[i]\backslash\{0\}$ and $v \in \mathcal{W}_{i-1}$, \label{enum:sequence-vector-division}
\end{enumerate}
where $\mathcal{W}_i = \mathcal{W}_{i-1} \cup \{w_i\}$. This procedure ends when the vectors $q_j$, for $j=0,\dots, 2k-2$, have been added. That is $q_j \in \mathcal{W}_N$ for $j=0,\dots,2k-2$ and some $N \in \mathbb{Z}_{\ge 0}$. This is akin to the definition of a linear circuit \cite{chazelle1998spectral, morgenstern1971linear, morgenstern1973note}. The reader may verify that the Gaussian elimination process may be captured in this manner.

In due course, we shall consider the $\mathbb{Z}[i]$-modules $\Lambda_i \subseteq W$ generated by $\mathcal{W}_i$. The $\mathbb{Z}[i]$-modules $\Lambda_i$ are free and of maximal rank $|S|=2k-1$, which are also referred to as $\mathbb{Z}[i]$-lattices. The final $\mathbb{Z}[i]$-lattice $\Lambda_N$ contains the $\mathbb{Z}[i]$-lattice generated by $q_j$ for $j=0,\dots,2k-2$, which we shall denote by $\Lambda$, that is
\begin{equation} \label{eq:division-final-lattice}
    \Lambda \coloneqq \Span_{\mathbb{Z}[i]}\left(\{ q_j \mid j=0,\dots,2k-2 \}\right) \subseteq \Lambda_N.
\end{equation}
As the sets $\mathcal{W}_i$ form an increasing sequence, so do the $\mathbb{Z}[i]$-lattices $\Lambda_i$:
\begin{equation} \label{eq:interpolation-lattices-sequence}
    \Lambda_0 \subseteq \Lambda_1 \subseteq \dots \subseteq \Lambda_N.
\end{equation}
The inclusion $\Lambda_{i-1} \subseteq \Lambda_{i}$, for $1\le i \le N$, is strict only if the newly added vector is of the second type \ref{enum:sequence-vector-division}, i.e., a division occurs. Furthermore, the relative index $[\Lambda_i:\Lambda_{i-1}]$ is a divisor of $\Nr(\lambda)$, where $\lambda$ is the divisor in the division \ref{enum:sequence-vector-division}. If we localise the sequence at a Gaussian prime $p\in \mathbb{Z}[i]$, we get an increasing sequence of $\mathbb{Z}[i]_p$-lattices
\begin{equation} \label{eq:interpolation-lattices-sequence-localised}
    \Lambda_{0,p} \subseteq \Lambda_{1,p} \subseteq \dots \subseteq \Lambda_{N,p},
\end{equation}
where the inclusions $\Lambda_{i-1,p} \subseteq \Lambda_{i,p}$, for $1\le i \le N$, are strict only if the corresponding division in \ref{enum:sequence-vector-division} is by a multiple of $p$.

\subsection{A simple lower bound} \label{sec:divisions-lower-bound-simple}

\begin{prop} \label{prop:division_by_p}
    Let $p\in \mathbb{Z}[i]$ be a Gaussian prime. If $2k-1 \ge \Nr(p) + 2$, then the interpolation phase \ref{enum:Gaussian_Toom_Cook_interpolation} of the Gaussian Toom--Cook-$k$ algorithm contains a division by a multiple of $p$.
\end{prop}

\begin{proof}
    It is sufficient to show that the integral Gaussian matrix associated to the system of linear equations in the interpolation phase, see Equation \eqref{eq:general-interpolation-linear-system}, has determinant divisible by $p$. Since said matrix is a Vandermonde matrix \eqref{eq:Vandermonde-abbreviation}, we can compute the determinant explicitly:
    \begin{equation} \label{eq:vandermonde}
        \det(V(S)) 
        = \epsilon \prod_{\substack{\{(x,y), (u,v)\} \subseteq S \\ (x,y) \neq (u,v)}} \left( xv-uy \right), \quad \text{where } \epsilon \in \{\pm 1\}.
    \end{equation}
    We may assume that for all $(x,y)\in S$ we have $(x,y) \nequiv (0,0) \mod(p)$ as otherwise the determinant \eqref{eq:vandermonde} is divisible by $p^{2k-2}$. Now, the determinant \eqref{eq:vandermonde} vanishes modulo $p$ if and only if there are two evaluation points in $S$ that project to the same point on the projective line $P^1(\mathbb{Z}[i] \slash p\mathbb{Z}[i])$. The latter, however, only has $|\mathbb{Z}[i] \slash p\mathbb{Z}[i]| + 1 = \Nr(p)+1$ points. Hence, if $|S| = 2k-1 \ge \Nr(p) + 2$ then two evaluation points in $S$ must project to the same point by the pigeonhole principle. Thus, the Vandermonde determinant \eqref{eq:vandermonde} is divisible by $p$.
\end{proof}

\begin{corollary} \label{cor:Toom--Cook-3+-contains-division}
    Every Gaussian Toom--Cook-$k$ algorithm with $k \ge 3$ contains at least a division (by a non-unit) in the interpolation phase \ref{enum:Gaussian_Toom_Cook_interpolation}.
\end{corollary}

\begin{proof}
    Apply Proposition \ref{prop:division_by_p} with the Gaussian prime $1+i$.
\end{proof}

\begin{remark} \label{rem:single-division-p}
    Although Proposition \ref{prop:division_by_p} is stated over the Gaussian integers, it holds more generally over integral domains $R$ with $\Nr(p)$ replaced by $| R\slash pR |$.
\end{remark}

\subsection{A linear lower bound} \label{sec:divisions-linear-lower-bound}

Theorem \ref{thm:main-divisions} is an immediate consequence of the following more general theorem.

\begin{theorem} \label{thm:divisions-gaussian-toom-cook-general}
    Let $p \in \mathbb{Z}[i]$ be a Gaussian prime. The number of division by multiples of $p$ in the interpolation phase \ref{enum:Gaussian_Toom_Cook_interpolation} of the Gaussian Toom--Cook-$k$ algorithm is at least $2k-2-\Nr(p)$.
\end{theorem}

In turn, Theorem \ref{thm:divisions-gaussian-toom-cook-general} is the combination of the following two propositions.

\begin{prop} \label{prop:number-division-at-least-divisible-elementary-divisors}
    Let $p \in \mathbb{Z}[i]$ be a Gaussian prime. The number of division by multiples of $p$ in the interpolation phase \ref{enum:Gaussian_Toom_Cook_interpolation} of the Gaussian Toom--Cook-$k$ algorithm is at least the number of elementary divisors of $V(S)$ divisible by $p$.
\end{prop}

\begin{prop} \label{prop:number-divisible-elementary-divisors}
    Let $p \in \mathbb{Z}[i]$ be a Gaussian prime and $V(S)$ the Vandermonde matrix \eqref{eq:Vandermonde-abbreviation} in the interpolation phase \ref{enum:Gaussian_Toom_Cook_interpolation}. Then, the number of elementary divisors of $V(S)$ divisible by $p$ is at least $2k-2-\Nr(p)$.
\end{prop}

\begin{remark} \label{rem:linear-divisions-p}
    Although Theorem \ref{thm:divisions-gaussian-toom-cook-general} and Propositions \ref{prop:number-division-at-least-divisible-elementary-divisors}, \ref{prop:number-divisible-elementary-divisors} are stated over the Gaussian integers, they hold more generally over principal ideal domains $R$ with $\Nr(p)$ replaced by $| R\slash pR |$.
\end{remark}


We begin with the proof of Proposition \ref{prop:number-divisible-elementary-divisors}.



\begin{proof}[Proof of Proposition \ref{prop:number-divisible-elementary-divisors}]
Let $e_1 \mid e_2 \mid \dots \mid e_{2k-1}$ be the elementary divisors of $V(S)$ and define for two equinumerous sets $T \subseteq S$ and $\mathcal{J} \subseteq \{0,1,\dots, 2k-2\}$:
\begin{equation} \label{eq:general-Vandermonde}
    V(T; \mathcal{J}) \coloneqq \begin{bmatrix} x^j y^{2k-2-j} \end{bmatrix}_{(x,y)\in T,\, j \in \mathcal{J}}.
\end{equation}
It is well known, see for example \cite[Satz 6, p.\@ 77]{Bosch-Alg}, that the product of the first $n$ elementary divisors is (up to multiplication by a unit) equal to the greatest common divisor of all determinants of $n \times n$ minorants. That is for $n = 1,\dots,2k-1$ there is a unit $\epsilon \in \mathbb{Z}[i]^{\times}$ such that
\begin{equation}
    e_1 \cdots e_n = \epsilon \cdot \gcd_{\substack{ T \subseteq S \\ \mathcal{J} \subseteq \{0,\dots, 2k-2\} \\ |T| = |\mathcal{J}| = n}} \det(V(T; \mathcal{J})).
\end{equation}
    It is well known and not hard to show, see for example \cite[Ch.\@ XI]{muir-treatise-determinants}, that
\begin{equation}
    \det(V(T)) \mid \det(V(T; \mathcal{J})).
\end{equation}
Hence, 
\begin{equation}
    \gcd_{\substack{ T \subseteq S \\ |T| = n}} \det(V(T)) \mid e_1 \cdots e_n \mid e_n^n.
\end{equation}
It now follows from the proof of Proposition \ref{prop:division_by_p} that $p \mid e_n$ for $n \ge \Nr(p) + 2$. This concludes the proof.

\end{proof}

In order to prove Proposition \ref{prop:number-division-at-least-divisible-elementary-divisors}, we require a further lemma, which should be known but we were unable to locate it in the literature.

\begin{lemma} \label{lem:elementary-divisors-of-chains-of-modules}
    Let $R$ be a principal ideal domain and $L \subseteq M \subseteq N$ be an increasing sequence of free $R$-left-submodules of equal rank $n$. Denote by $e_i(L:M)$ the $i$-th elementary divisor of $L \subseteq M$ and analogously for the inclusions $M \subseteq N$ and $L \subseteq N$. Then, we have for $i=1,\dots,n$ that
    \begin{equation}
        \prod_{j=1}^i e_j(L:M) \cdot \prod_{j=1}^i e_j(M:N) \, \Big | \, \prod_{j=1}^i e_j(L:N). 
    \end{equation}
\end{lemma}

\begin{proof}
    Let us first consider the case $i=1$. Fix bases 
    of $L$, $M$, and $N$, respectively. Let $C$ be the matrix expressing the basis of $L$ in terms of the basis of $N$, $A$ be the matrix expressing the basis of $L$ in terms of the basis of $M$, and $B$ be the matrix expressing the basis of $M$ in terms of the basis of $N$. Then, the first elementary divisor $e_1(L:N)$ of $L \subseteq N $ is a greatest common divisor of the coefficients of $C$ and analogously for $e_1(L:M)$ and $e_1(M:N)$. Since $C=AB$, it follows that $e_1(L:M) e_1(M:N) \mid e_1(L:N)$.

    For the general case, we take the $i$-th exterior product and consider the increasing sequence $\bigwedge^i L \subseteq \bigwedge^i M \subseteq \bigwedge^i N$. Now, the product $\prod_{j = 1}^i e_j(L:N)$ of elementary divisors of $L \subseteq N$ is the first elementary divisor of $\bigwedge^i L \subseteq \bigwedge^i N$ and analogously for the other inclusions (cf.\@ \cite[proof of Satz 6, pp.\@ 77-79]{Bosch-Alg}). The claim thus follows from the previously considered case.
\end{proof}

We are now in a position to prove Proposition \ref{prop:number-division-at-least-divisible-elementary-divisors}.

\begin{proof}[Proof of Proposition \ref{prop:number-division-at-least-divisible-elementary-divisors}]
The elementary divisors of $V(S)$ are (by definition) equal to the elementary divisors of $\Lambda_0 \subseteq \Lambda$. Since $\Lambda_0 \subseteq \Lambda \subseteq \Lambda_N$, we have that the product of the first $n$ elementary divisors of $\Lambda_0 \subseteq \Lambda$ divides the product of the first $n$ elementary divisors of $\Lambda_0 \subseteq \Lambda_N$ for all $n$ by Lemma \ref{lem:elementary-divisors-of-chains-of-modules}. In particular, the number of elementary divisors of $\Lambda_0 \subseteq \Lambda_N$ divisible by $p$ is at least the number of elementary divisors of $\Lambda_0 \subseteq \Lambda$ divisible by $p$. Denote by $e_1 \mid e_2 \mid \dots \mid e_{2k-1}$ the elementary divisors of $\Lambda_0 \subseteq \Lambda_N$ and by $\alpha$ the number of those which are divisible by $p$. Then, we have an isomorphism of $\mathbb{Z}[i]$-modules:
\begin{equation}
    \Lambda_N \slash \Lambda_0 \cong \left( \mathbb{Z}[i] \slash e_1 \mathbb{Z}[i] \right) \times \left( \mathbb{Z}[i] \slash e_2 \mathbb{Z}[i] \right) \times \dots \times \left( \mathbb{Z}[i] \slash e_{2k-1} \mathbb{Z}[i] \right).
\end{equation}
We may then look at the quotient
\begin{equation}
    \rquotient{\Lambda_N \slash \Lambda_0}{p(\Lambda_N \slash \Lambda_0)} \cong \left( \mathbb{Z}[i] \slash p \mathbb{Z}[i] \right)^{\alpha} \cong \mathbb{F}_{\Nr(p)}^{\alpha}
\end{equation}
as a $\mathbb{Z}[i] \slash p \mathbb{Z}[i] \cong \mathbb{F}_{\Nr(p)}$ vector space, which has dimension equal to $\alpha$.
Let us now localise at $p$. We have similarly
\begin{equation}
    \Lambda_{N,p} \slash \Lambda_{0,p} \cong \left( \mathbb{Z}[i] \slash p^{\alpha_1} \mathbb{Z}[i] \right) \times \left( \mathbb{Z}[i] \slash p^{\alpha_2} \mathbb{Z}[i] \right) \times \dots \times \left( \mathbb{Z}[i] \slash p^{\alpha_{2k-1}} \mathbb{Z}[i] \right),
\end{equation}
where $\alpha_i$ is the exponent of $p$ in the Gaussian prime factorisation of $e_i$, and 
\begin{equation}
    \rquotient{\Lambda_{N, p} \slash \Lambda_{0,p}}{p(\Lambda_{N,p} \slash \Lambda_{0,p})} \cong \left( \mathbb{Z}[i] \slash p \mathbb{Z}[i] \right)^{\alpha} \cong \mathbb{F}_{\Nr(p)}^{\alpha}.
\end{equation}
%
It follows that the minimal number of generators of the $\mathbb{Z}[i]_p$-module $\Lambda_{N,p} \slash \Lambda_{0,p}$ is at least $\alpha$. On the other hand, by looking at the increasing sequence
\begin{equation}
    \Lambda_{0,p} \slash \Lambda_{0,p} \subseteq \Lambda_{1,p} \slash \Lambda_{0,p} \subseteq \dots \subseteq \Lambda_{N,p} \slash \Lambda_{0,p},
\end{equation}
a single new linear independent element may only be added in the sequence if and only if the inclusion $\Lambda_{i-1,p} \slash \Lambda_{0,p} \subsetneq \Lambda_{i,p} \slash \Lambda_{0,p}$ is strict. This may only occur if a division by a multiple of $p$ takes place in \ref{enum:sequence-vector-division}, cf.\@ Equation \eqref{eq:interpolation-lattices-sequence-localised} and surrounding remarks. The proof is thus concluded.
\end{proof}

\subsection{An upper bound}

In this section, we prove the existence of a Gaussian Toom--Cook-$k$ algorithm which matches the lower bound on the number of non-trivial divisions in Theorem~\ref{thm:main-divisions}.

\begin{prop}
    For every $k \ge 2$, there is a Gaussian Toom--Cook-$k$ algorithm which only uses $\max\{0,2k-7\}$ non-trivial divisions.
\end{prop}
\begin{proof}
    We may restrict ourselves to $k \ge 6$ as we have already discussed the smaller cases. By the elementary divisor theorem, we may write
    \begin{equation}
    \Lambda \slash \Lambda_0 \cong \left( \mathbb{Z}[i] \slash e_1 \mathbb{Z}[i] \right) \times \left( \mathbb{Z}[i] \slash e_2 \mathbb{Z}[i] \right) \times \dots \times \left( \mathbb{Z}[i] \slash e_{2k-1} \mathbb{Z}[i] \right).
\end{equation}
    This means, we can perform a base change in $\Lambda_0$, then divide by $e_i$ the corresponding vector for $i=1,\dots,2k-1$, followed by a base change in $\Lambda$ to get the elements we desire. Both base changes involve no divisions as they are just a multiplication by a matrix in $\GL_{2k-1}(\mathbb{Z}[i])$. Thus, this algorithm uses $2k-1$ divisions \emph{a priori}. To get to $2k-7$ divisions, we are left to show that there are evaluation points such that the elementary divisors $e_1, e_2, \dots, e_6$ are powers of $(1+i)$ times a unit as divisions by $(1+i)$ can be realised as a multiplication by $(1-i)$ and a division by $2$. Due to the nature of elementary divisors being given by greatest common divisors of determinants of minors, it suffices to find a minor with said property. Several minors are given by Vandermonde matrices \eqref{eq:general-Vandermonde} of subsets of the evaluation points. Thus, it suffices to find six points such that the corresponding Vandermonde matrix has the desired property. We choose the points
    \begin{equation}
        S = \left\{ (0,1), (1,1), (-1, 1), (i, 1), (-i, 1), (1, 0) \right\},
    \end{equation}
    whose Vandermonde matrix has the elementary divisors
    $
        1 \mid 1 \mid 1 \mid (1+i) \mid (1+i)^3 \mid (1+i)^4
    $,
    which are easily computed using Sage~\cite{sagemath}.
\end{proof}

\section{The division factor of Gaussian Toom--Cook algorithms}
\label{sec:division-factor}

The division factor of a Toom--Cook-$k$ algorithm is the smallest positive rational integer $\alpha \in \mathbb{Z}_{>0}$ such that there exists an equivalent Toom--Cook-$k$ algorithm involving only divisions by a power of $2$ to compute $\alpha$-times the product of two (projective) polynomials of degree $k-1$, cf.\@ \cite[Def.~2.4]{nissim2025minimizing}. We adopt the same definition over the Gaussian integers. We are able to prove the following lower bound on the division factor of Gaussian Toom--Cook-$k$ algorithms.

\begin{theorem}\label{thm:division-factor}
    Denote by $\mathcal{P} \subseteq \mathbb{Z}[i]$ a representative set of all Gaussian primes up to association (multiplication by unit). The division factor of any Gaussian Toom--Cook-$k$ algorithm is divisible by
    \begin{equation}
         \prod_{\substack{p \in \mathcal{P} \\ p\, \nmid \,2}} p^{\left\lfloor \frac{2k-2}{\Nr(p)+1} \right\rfloor}.
    \end{equation}
\end{theorem}

\begin{proof}
    We shall follow the proof strategy of \cite{nissim2025minimizing}. We prove this by showing that the least common multiple of the denominators of the interpolation linear transformation is divisible by $p^{\lfloor \frac{2k-2}{\Nr(p)+1} \rfloor}$ for every Gaussian prime $p$. We remark that the reciprocal of this value, which is equal to the greatest common divisor (as a fractional ideal) of the image of integral vectors is often called the \emph{content} of a linear map.

    The inverse of the interpolation linear transformation is given by the linear map which evaluates the polynomials in $\mathbb{Z}[i][X,Y]$ of homogeneous degree $2k-2$ at the evaluation points $S \subset \mathbb{Z}[i]^2 \backslash\{(0,0)\}$:
    \begin{equation} \begin{aligned}
        W^{-T}: \mathbb{Z}[i][X,Y]_{2k-2} &\to \mathbb{Z}[i]^S \\
        P(X,Y) & \mapsto [P(x,y)]_{(x,y)\in S}.
    \end{aligned}
    \end{equation}
    Let $p \in \mathbb{Z}[i]$ be a Gaussian prime and abbreviate
    \begin{equation}
        \ell = \left \lfloor \frac{2k-2}{\Nr(p)+1} \right \rfloor.
    \end{equation}
    Consider the polynomial 
    \begin{equation}
    R(X,Y) = (X^{\Nr(p)}Y-XY^{\Nr(p)})^{\ell} Y^{2k-2-(\Nr(p)+1)\ell}.
    \end{equation}
    It is of homogeneous degree $2k-2$ and represents a primitive vector, that is there is no non-unit $\beta \in \mathbb{Z}[i]\backslash\{0\}$ such that $\beta^{-1}R(X,Y) \in \mathbb{Z}[i][X,Y]_{2k-2}$. Indeed, the coefficient of the monomial $X^{\Nr(p)\ell} Y^{2k-2-\Nr(p)\ell}$ in $R(X,Y)$ is $1$. On the other hand, we have
    \begin{equation}
    x^{\Nr(p)} \equiv x, \mod(p), \quad \forall x \in \mathbb{Z}[i],
    \end{equation}
    by Fermat's little theorem, and hence
    \begin{equation}
        p^{\ell} \mid R(x,y), \quad \forall (x,y) \in S.
    \end{equation}
    It follows that the image of $R(X,Y)$ under the evaluation linear map $W^{-T}$ is integrally divisible by $p^{\ell}$. Thus, the least common multiple of the denominators of the interpolation linear map $W^T$ must be divisible by $p^{\ell}$ as it is the inverse of the evaluation linear map $W^{-T}$. Indeed, if this were not the case, we would have
    \begin{equation}
        R(X,Y) = W^T W^{-T} R(X,Y) \in W^T p^{\ell} \mathbb{Z}[i]^S \subseteq p (\mathbb{Z}[i][X,Y]_{2k-2})_{p},
    \end{equation}
    where subscript $p$ denotes the localisation at the prime $p$. This is a contradiction as $R(X,Y)$ is a primitive vector in $\mathbb{Z}[i][X,Y]_{2k-2}$ and hence also in $(\mathbb{Z}[i][X,Y]_{2k-2})_{p}$. We conclude the theorem after taking the least common multiple over all primes which do not divide $2$.
\end{proof}

	\section{Empirical evaluation}
\label{sec:bench}

\subsection{Implementation}

Assisted by AI, we have implemented our rational-to-Gaussian and Gaussian Toom--Cook algorithms in C and assembly as an extension to the \emph{GNU Multiple Precision Arithmetic Library} (GMP)~\cite{gmp}. It is available on Github~\cite{libgtc2026}.

In order to get the most performance out of our algorithms, we employed many optimisation strategies present in GMP and other high-performance-computing libraries. This includes for example allocating a memory arena to minimise allocations, operation fusing and instruction reordering for improved instruction level parallelism. During our optimisation efforts, we have found three places where a slight alteration of our algorithm led to increased performance:

\begin{enumerate}
	\item In the rational-to-Gaussian Toom--Cook reduction algorithms, it is better to use the evaluation point $(-1,1)$ instead of $(1,1)$. The reason is that this decreases the likelihood of an extra carry limb since GMP and our extension library operate on unsigned integers.
	\item In Gaussian Toom--Cook-$4$, for the evaluation of the point $(2,1)$, it is faster to use Horner's method~\cite{horner1819new} despite an extra arithmetic operation. The reason is a combination of reduced memory access and operation fusing which arithmetic complexity does not take into account.
	\item In the butterfly opening of Gaussian Toom--Cook-$5$, it is faster to do a fused (early) division by $4$ and then remultiply by $4$ when needed even though this costs an extra operation. The reason is again reduced memory access and operator fusing.
\end{enumerate}

Whilst the latter two are a minor change in our algorithms, the former is easily accommodated in the evaluation and interpolation schemes by changing the sign of all finite points.

Analogous to GMP, we allowed our Gaussian Toom--Cook algorithms to recursively call Gaussian Toom--Cook algorithms of equal or lower order (including a Gaussian Karatsuba algorithm) as well as a Gaussian textbook algorithm. Our rational multiplication algorithms, rational-to-Gaussian Toom--Cook, are allowed to make similar recursive calls. For recursive calls invoking Gaussian multiplication, it is allowed to call our Gaussian Toom--Cook algorithms of equal or lower order (including a Gaussian Karatsuba algorithm) and the Gaussian textbook algorithm. For recursive calls invoking rational multiplication, it is allowed to call rational-to-Gaussian Toom--Cook algorithms of equal and lower order as well as a rational textbook algorithm and GMP rational Karatsuba algorithm. We decided to include the latter as the rational-to-Gaussian Karatsuba algorithm combined with Gaussian Karatsuba shares many similarities with the rational Karatsuba algorithm.

\subsection{Experimental setup}

We compare our rational and Gaussian multiplication algorithms against their equivalent counterpart in the \emph{GNU Multiple Precision Arithmetic Library} (GMP) version 6.3.0~\cite{gmp}.

For rational multiplication algorithms, we compare our rational-to-Gaussian Toom--Cook-$k$ algorithm against GMP's Toom--Cook-$k$ algorithm. We note that GMP does not have a Toom--Cook-$5$ algorithm. Thus, in this case, we compare against GMP Toom--Cook-$4$ and Toom--Cook-$6.5$ algorithm. The latter short-circuits to an internal Toom--Cook-$6$ algorithm when the numbers are (nearly) balanced (as is the case in our evaluation).

For Gaussian multiplication algorithms, we compare our Gaussian Toom--Cook-$k$ algorithms against a three rational multiplication algorithm using GMP's Toom--Cook-$k$ for the rational multiplications. Again, the same comment about Toom--Cook-$5$ applies here.

We compare the Toom--Cook-$k$ algorithms on products of equal size up to $k \cdot 1000$ limbs, which are the basic multiplication block size of the architecture. This covers the relevant ranges for each algorithm \cite{bodrato}. We carry out these experiments on three architectures covering both x86 and ARM. The concrete specifications for the three machines are as follows:

\begin{itemize}
	\item Huawei Kunpeng 920-4826 (Hi1620) processor (ARM), with 512 GB memory, theoretical peak memory throughput of 187.7 GB/s, and two sockets with 48 cores each for a total of 96 cores; 2 NUMA domains per socket; kernel version 5.15.0; GCC version 11.4.0.
	\item AMD EPYC 7763 processor (x86), with 1024 GB memory, theoretical peak memory throughput of 204.8 GB/s, and two sockets with 64 cores each for a total of 128 cores; 1 NUMA domain per socket; kernel version 5.15.0; GCC version 11.4.0;
	\item Intel Xeon Gold 6238T processor (x86), with 192 GB memory, theoretical peak memory throughput of 140.8 GB/s, and two sockets with 22 cores each for a total of 44 cores; 1 NUMA domain per socket; kernel version 5.14.0; GCC version 11.5.0.
\end{itemize}

\subsection{Results}

\subsubsection{Rational multiplication}

\begin{figure}[htpb]
	\centering
	\includegraphics[width=0.95\linewidth]{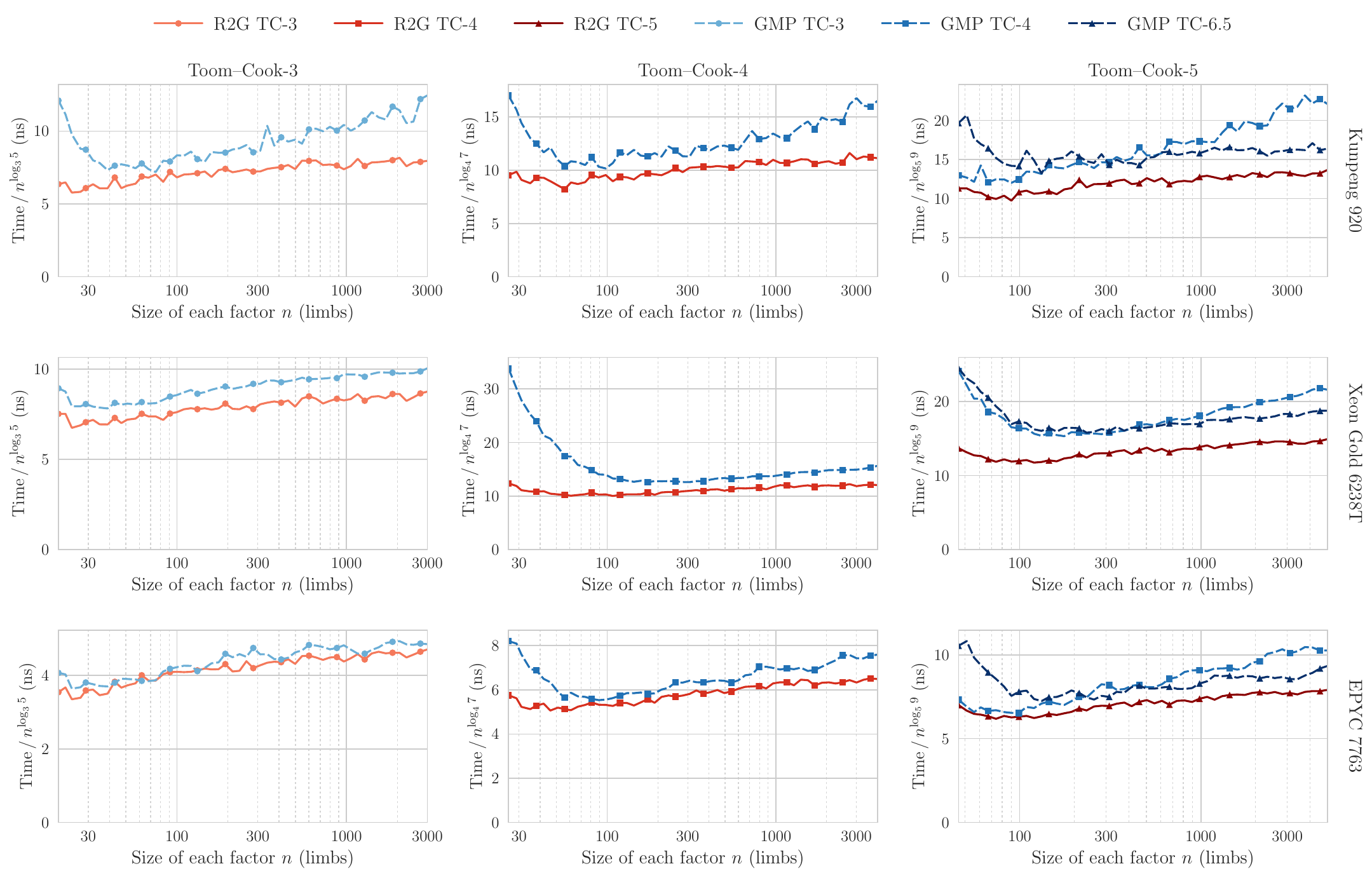}
	\caption{Normalised runtime of rational multiplication algorithms on products of two rational integers with $n$ limbs on several architectures. The runtime in nano seconds (ns) is normalised with respect to the asymptotic complexity $n^{\log_k (2k-1)}$ of the Toom--Cook-$k$ algorithms. Displayed are our rational-to-Gaussian Toom--Cook algorithms (R2G TC) as well as GMP's Toom--Cook algorithms (GMP TC).}
	\label{fig:rational-bench}
\end{figure}

Figure~\ref{fig:rational-bench} compares our rational multiplication algorithms, rational-to-Gaussian Toom--Cook-$k$ (R2G TC-$k$), with the corresponding Toom--Cook-$k$ algorithm in GMP (GMP TC-$k$). Over relevant ranges, we find the following speed-ups:

\begin{itemize}
	\item R2G TC-3 displays an 18\%, 13\%, respectively 4\%, geometric-mean speed increase over GMP TC-3 across limbs in the range $[35, 300[$ on the Kunpeng, Xeon Gold, respectively EPYC, machine;
	\item R2G TC-4 displays a 19\%, 29\%, respectively 8\%, geometric-mean speed increase over GMP TC-4 across limbs in the range $[60, 350[$ on the Kunpeng, Xeon Gold, respectively EPYC, machine;
	\item R2G TC-5 displays a 24\%, 29\%, respectively 11\%, geometric-mean speed increase over GMP TC-4 across limbs in the range $[70, 500[$ on the Kunpeng, Xeon Gold, respectively EPYC, machine;
	\item R2G TC-5 displays a 29\%, 29\%, respectively 15\%, geometric-mean speed increase over GMP TC-6.5 across limbs in the range $[70, 3000[$ on the Kunpeng, Xeon Gold, respectively EPYC, machine.
\end{itemize}

\subsubsection{Gaussian multiplication}

\begin{figure}[htpb]
	\centering
	\includegraphics[width=0.95\linewidth]{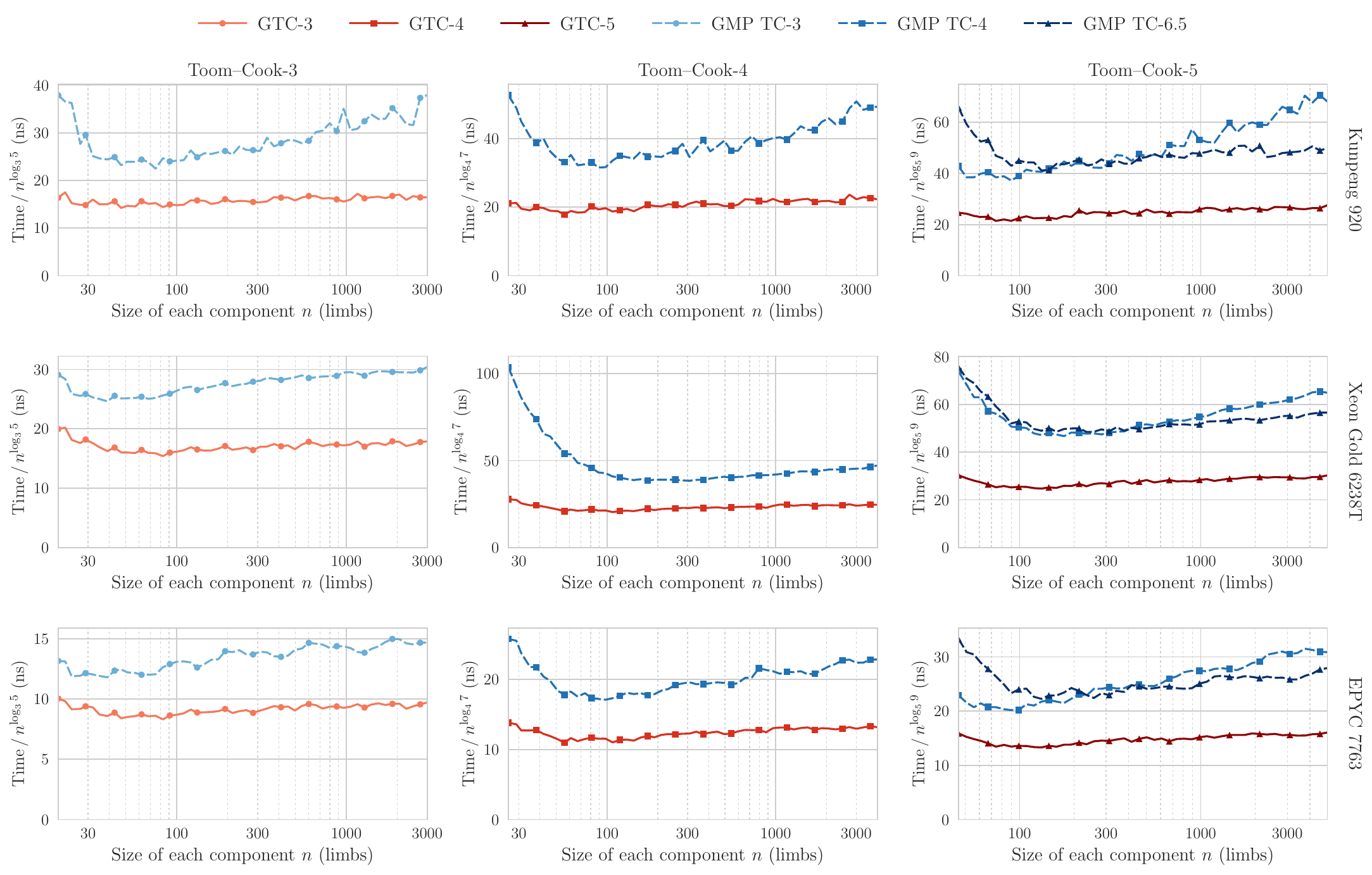}
	\caption{Normalised runtime of Gaussian multiplication algorithms on products of two Gaussian integers with $n$ limbs on several architectures. The runtime in nano seconds (ns) is normalised with respect to the asymptotic complexity $n^{\log_k (2k-1)}$ of the Toom--Cook-$k$ algorithms. Displayed are our Gaussian Toom--Cook algorithms (GTC) as well as a 3 rational multiplication algorithm using GMP's Toom--Cook algorithms (GMP TC).}
	\label{fig:Gauss-bench}
\end{figure}

Figure~\ref{fig:Gauss-bench} compares our Gaussian multiplication algorithms, Gaussian Toom--Cook-$k$ (GTC-$k$), with a three rational multiplication algorithm using the corresponding Toom--Cook-$k$ algorithm from GMP (GMP TC-$k$). Over relevant ranges, we find the following speed-ups:

\begin{itemize}
	\item GTC-3 displays a 63\%, 61\%, respectively 47\%, geometric-mean speed increase over GMP TC-3 across limbs in the range $[40, 300[$ on the Kunpeng, Xeon Gold, respectively EPYC, machine;
	\item GTC-4 displays a 74\%, 86\%, respectively 55\%, geometric-mean speed increase over GMP TC-4 across limbs in the range $[60, 500[$ on the Kunpeng, Xeon Gold, respectively EPYC, machine;
	\item GTC-5 displays an 81\%, 87\%, respectively 61\%, geometric-mean speed increase over GMP TC-4 across limbs in the range $[80, 500[$ on the Kunpeng, Xeon Gold, respectively EPYC, machine;
	\item GTC-5 displays an 87\%, 88\%, respectively 67\%, geometric-mean speed increase over GMP TC-6.5 across limbs in the range $[80, 3000[$ on the Kunpeng, Xeon Gold, respectively EPYC, machine.
\end{itemize}

	\section{Discussion}\label{sec:discussion}

In this work, we challenged the conventional hierarchy of algebraic complexity by demonstrating that long rational integer multiplication can be computed more efficiently by reducing it to long Gaussian integer multiplication. By shifting the evaluation and interpolation sequences to the Gaussian field $\mathbb{Q}(i)$, our Toom--Cook reduction leverages denser interpolation points and algebraic symmetries unavailable in the rational field $\mathbb{Q}$. This structural shift not only reduces the overall arithmetic cost but also decisively circumvents established theoretical lower bounds for rational Toom--Cook algorithms regarding division factors and non-trivial divisions \cite{nissim2025minimizing, Toom-Cook-divisions}. The theoretical gains directly translate into practice, as evidenced by our implementation outperforming the highly optimised GNU Multiple Precision Arithmetic Library (GMP) by up to 29\%, establishing a new practical standard for integer multiplication in this regime.

While our focus has been on Toom--Cook algorithms with arithmetic complexity $O(n^{\log_k(2k-1)})$ for small values of $k$ \cite{karatsuba-multiplication, toom, cook}, this inverted reduction paradigm opens the door to analogous improvements in asymptotically faster algorithms. A natural direction for future research is the development of a Gaussian Schönhage--Strassen algorithm. Fast Fourier Transform (FFT) based integer multiplication algorithms already achieve $O(n \operatorname{polylog}(n))$ arithmetic complexity \cite{schonhage, furer2009faster, harvey2016even, harvey2021integer, OAI:Integer-multiplication-below-n-log-n-September-23-2026}, but they are often burdened by large constant factors that render them impractical for mid-range inputs. Adapting our Gaussian reduction techniques to FFT-based models could theoretically reduce these hidden constants by exploiting the fourth roots of unity and eighth roots of 16 inherent to the Gaussian integers, effectively optimising the underlying convolution structures.

Beyond asymptotic improvements, our methodology has broader implications for specialised computational environments. Since long integer multiplication is a critical bottleneck in various domains, our Gaussian Toom--Cook matrices could be adapted for algorithm-based fault tolerance in parallel computation frameworks, where minimising exact multi-precision divisions is essential for robust and efficient error correction. Furthermore, reducing the division factor and eliminating non-trivial divisions is highly advantageous for quantum computing circuits, where exact divisions heavily dominate Toffoli depth and count \cite{gu2018division, larasati2021quantum,putranto2023space}. Finally, implementing our Gaussian reduction scheme on hardware accelerators like FPGAs could yield significant area and delay trade-offs compared to traditional rational Toom--Cook multipliers \cite{das2024area}.

Another theoretical direction is the exploration of alternative ring extensions beyond the Gaussian integers. While $\mathbb{Z}[i]$ provides access to the fourth roots of unity and simplifies specific arithmetic operations, other algebraic number rings, such as the Eisenstein integers $\mathbb{Z}[\omega]$ (where $\omega = e^{2\pi i / 3}$), offer even richer symmetries. The Eisenstein integers contain the sixth roots of unity, which could provide a denser set of easily computable evaluation points for higher-order Toom--Cook parameterisations. Investigating general cyclotomic rings $\mathbb{Z}[\zeta_m]$ or higher-degree algebraic extensions could ostensibly reveal structural domains where the division factor is further minimised. 

However, scaling this framework to arbitrary ring extensions presents severe computational bottlenecks, rendering these directions largely unpromising for practical performance gains. The fundamental issue lies in the arithmetic overhead of the extension itself. In the Gaussian integers, the algebraic extension degree is only two, meaning the constant-factor cost of ring additions and multiplications remains small enough to be entirely eclipsed by the algorithmic savings of the Toom--Cook reduction.

Regarding the algorithmic savings, finding the optimal number of operations needed for computing a linear sequence is NP-hard and APX-complete. Traditional search algorithms find near-optimal solutions, and the technique introduced by Bodrato and Zanoni \cite{bodrato} is an example of that. The problem with search algorithms is that they get stuck in a local minimum solution. But two solutions that are considered far away from a search algorithm's perspective may seem close from a human's or AI's perspective. Our approach to escape local minima incorporates AI to iteratively find a close better solution (close from AI's perspective) that escapes the local minimum (from the search algorithm's perspective), and then feeds it back to the search algorithm. Using this technique we managed to further reduce the number of operations needed for our Gaussian Toom--Cook-$5$ algorithm. We expect this approach can be incorporated to obtain better solutions for other search problems as well.

On the practical side, the implementation of Gaussian integer arithmetic has demonstrated further advantages, but also disadvantages. One disadvantage that showed up is that Gaussian integers have twice the amount of carry limbs: once for the real part and once for the imaginary part. This has a negative impact for small limb multiplication. On the other hand, on modern architectures, textbook Gaussian multiplication executes much more efficiently than traditionally assumed. This is thanks to better utilisation of instruction-level parallelism. Direct Gaussian integer multiplication is only $2\times$ to $2.5\times$ slower than its rational equivalent, as opposed to the $6\times$ assumption in our cost model.

Overall, we have demonstrated that Toom--Cook algorithms for rational integer multiplication can be significantly enhanced by leveraging their Gaussian integer counterparts. However, the Toom--Cook family is not the only class of long integer multiplication algorithms used in practice. We anticipate that a similar Gaussian-domain approach could yield comparable improvements for the Schönhage--Strassen algorithm, unbalanced integer multiplication algorithms, and broader applications in general integer and polynomial multiplication.

	\bibliography{references}

\newcommand{\etalchar}[1]{$^{#1}$}
\begin{thebibliography}{HVDHL16}

\bibitem[AAA20]{ahmed2020atm}
Mohammed~Salman Ahmed, Deepthi Amuru, and Zia Abbas.
\newblock {ATM}: approximate {T}oom--{C}ook multiplication for speech
  processing applications.
\newblock In {\em 2020 IEEE International Symposium on Circuits and Systems
  (ISCAS)}, pages 1--5. IEEE, 2020.

\bibitem[AM69]{atiyah-macdonald-commutative-algebra}
Michael~F. Atiyah and Ian~G. Macdonald.
\newblock {\em Introduction to commutative algebra}.
\newblock Reading, {Mass}.-{Menlo} {Park}, {Calif}.-{London}-{Don} {Mills},
  {Ont}.: {Addison}-{Wesley} {Publishing} {Company}, 1969.

\bibitem[AMR23]{allam2023comparative}
Harish~Prasad Allam, Suraj Mandal, and Debapriya~Basu Roy.
\newblock A comparative analysis between {K}aratsuba, {T}oom--{C}ook and {NTT}
  multiplier for polynomial multiplication in {NTRU} on {FPGA}.
\newblock In {\em 2023 Asian Hardware Oriented Security and Trust Symposium
  (AsianHOST)}, pages 1--6. IEEE, 2023.

\bibitem[BDS19]{bilardi2019complexity}
Gianfranco Bilardi and Lorenzo De~Stefani.
\newblock {{The {I/O} complexity of {T}oom--{C}ook integer multiplication}}.
\newblock In {\em Proceedings of the Thirtieth Annual ACM-SIAM Symposium on
  Discrete Algorithms}, pages 2034--2052. SIAM, 2019.

\bibitem[BM81]{brooks1981dynamics}
Robert Brooks and J~Peter Matelski.
\newblock The dynamics of 2-generator subgroups of {${\rm PSL}(2,
  \mathbb{C})$}.
\newblock In {\em Riemann surfaces and related topics: Proceedings of the 1978
  Stony Brook Conference, Ann. of Math. Stud}, volume~97, pages 65--71, 1981.

\bibitem[BMP08]{boyar-matthews-peralta-shortest-linear-program}
Joan Boyar, Philip Matthews, and Ren{\'e} Peralta.
\newblock On the shortest linear straight-line program for computing linear
  forms.
\newblock In Edward Ochma{\'{n}}ski and Jerzy Tyszkiewicz, editors, {\em
  Mathematical Foundations of Computer Science 2008}, pages 168--179, Berlin,
  Heidelberg, 2008. Springer Berlin Heidelberg.

\bibitem[BMP13]{BMP13}
Joan Boyar, Philip Matthews, and Ren\'{e} Peralta.
\newblock Logic minimization techniques with applications to cryptology.
\newblock {\em Journal of Cryptology}, 26(2):280--312, 2013.

\bibitem[Bod07]{bodrato2007notes}
Marco Bodrato.
\newblock Notes on low degree {T}oom--{C}ook multiplication with small
  characteristic.
\newblock {\em Preprint, Centro "V. Volterra", Universita di Roma "Tor
  Vergata"}, 2007.

\bibitem[Bos09]{Bosch-Alg}
Siegfried Bosch.
\newblock {\em Algebra}.
\newblock Springer-Lehrb. Berlin: Springer, 7th revised edition, 2009.

\bibitem[BS19]{beniamini}
Gal Beniamini and Oded Schwartz.
\newblock Faster matrix multiplication via sparse decomposition.
\newblock In {\em The 31st ACM Symposium on Parallelism in Algorithms and
  Architectures}, pages 11--22, 2019.

\bibitem[BZ07]{bodrato}
Marco Bodrato and Alberto Zanoni.
\newblock Integer and polynomial multiplication: towards optimal {T}oom--{C}ook
  matrices.
\newblock In {\em Proceedings of the 2007 international symposium on Symbolic
  and algebraic computation}, pages 17--24. ACM, 2007.

\bibitem[BZ10]{brent2010modern}
Richard~P. Brent and Paul Zimmermann.
\newblock {\em Modern computer arithmetic}, volume~18.
\newblock Cambridge University Press, 2010.

\bibitem[CA69]{cook}
Stephen~A. Cook and St{\aa}l~O. Aanderaa.
\newblock On the minimum computation time of functions.
\newblock {\em Transactions of the American Mathematical Society},
  142:291--314, 1969.

\bibitem[Cav73]{caviness1973lehmer}
Bob~F. Caviness.
\newblock A {L}ehmer-type greatest common divisor algorithm for {G}aussian
  integers.
\newblock {\em SIAM Rev}, 15(2):414, 1973.

\bibitem[CC76]{caviness1976algorithms}
Bob~F. Caviness and George~E. Collins.
\newblock Algorithms for {G}aussian integer arithmetic.
\newblock In {\em Proceedings of the third ACM symposium on Symbolic and
  algebraic computation}, pages 36--45, 1976.

\bibitem[CH07]{chung2007asymmetric}
Jaewook Chung and M.~Anwar Hasan.
\newblock Asymmetric squaring formulae.
\newblock In {\em 18th IEEE Symposium on Computer Arithmetic (ARITH'07)}, pages
  113--122. IEEE, 2007.

\bibitem[Cha98]{chazelle1998spectral}
Bernard Chazelle.
\newblock A spectral approach to lower bounds with applications to geometric
  searching.
\newblock {\em SIAM Journal on Computing}, 27(2):545--556, 1998.

\bibitem[Coo66]{cook-phd-thesis}
Stephen~A. Cook.
\newblock {\em On the minimum computation time of functions}.
\newblock PhD thesis, Harvard University, 1966.

\bibitem[CT65]{cooley1965algorithm}
James~W. Cooley and John~W. Tukey.
\newblock An algorithm for the machine calculation of complex {F}ourier series.
\newblock {\em Mathematics of computation}, 19(90):297--301, 1965.

\bibitem[DBC18]{dutta2018quantum}
Srijit Dutta, Debjyoti Bhattacharjee, and Anupam Chattopadhyay.
\newblock {{Quantum circuits for {T}oom--{C}ook multiplication}}.
\newblock {\em Physical Review A}, 98(1):012311, 2018.

\bibitem[DC85]{Davis1985}
Philip~J. Davis and William~G. Chinn.
\newblock {\em Pompeiu's Magic Seven}, pages 14--19.
\newblock Birkh{\"a}user Boston, Boston, MA, 1985.

\bibitem[DJ24]{das2024area}
Monalisa Das and Babita Jajodia.
\newblock Area and delay trade-offs in three-way {T}oom--{C}ook large integer
  multipliers implemented on {FPGA}s.
\newblock {\em IEEE Transactions on Circuits and Systems I: Regular Papers},
  2024.

\bibitem[DLG18]{ding2018high}
Jinnan Ding, Shuguo Li, and Zhen Gu.
\newblock {{High-speed ECC processor over NIST prime fields applied with
  {T}oom--{C}ook multiplication}}.
\newblock {\em IEEE Transactions on Circuits and Systems I: Regular Papers},
  66(3):1003--1016, 2018.

\bibitem[DS22]{de2022brief}
Lorenzo De~Stefani.
\newblock Brief announcement: on the {I/O} complexity of sequential and
  parallel hybrid integer multiplication algorithms.
\newblock In {\em Proceedings of the 34th ACM Symposium on Parallelism in
  Algorithms and Architectures}, pages 449--452, 2022.

\bibitem[DV90]{duhamel1990fast}
Pierre Duhamel and Martin Vetterli.
\newblock Fast {F}ourier transforms: a tutorial review and a state of the art.
\newblock {\em Signal processing}, 19(4):259--299, 1990.

\bibitem[Eli21]{elia2021loss}
Marcus Elia.
\newblock {\em Loss of precision in implementations of the {T}oom--{C}ook
  algorithm}.
\newblock The University of Vermont and State Agricultural College, 2021.

\bibitem[EW11]{emmart2011high}
Niall Emmart and Charles Weems.
\newblock High precision integer multiplication with a {GPU}.
\newblock In {\em 2011 IEEE International Symposium on Parallel and Distributed
  Processing Workshops and Phd Forum}, pages 1781--1787. IEEE, 2011.

\bibitem[Fog22]{Fog-InstructionTables}
Agner Fog.
\newblock Instruction tables.
\newblock \url{https://www.agner.org/optimize/instruction_tables.pdf}, 2022.
\newblock Accessed: 2025-06-17.

\bibitem[FSGL10]{fan2010overlap}
Haining Fan, Jiaguang Sun, Ming Gu, and K.-Y. Lam.
\newblock {{Overlap-free {K}aratsuba--{O}fman polynomial multiplication
  algorithms}}.
\newblock {\em IET Information security}, 4(1):8--14, 2010.

\bibitem[FSK10]{FS10}
Carsten Fuhs and Peter Schneider-Kamp.
\newblock Synthesizing shortest linear straight-line programs over {GF(2)}
  using {SAT}.
\newblock In {\em Theory and Applications of Satisfiability Testing ({SAT}
  2010)}, pages 71--84, 2010.

\bibitem[F{\"u}r09]{furer2009faster}
Martin F{\"u}rer.
\newblock Faster integer multiplication.
\newblock {\em SIAM Journal on Computing}, 39(3):979--1005, 2009.

\bibitem[Gil84]{gilbert1984arithmetic}
William~J. Gilbert.
\newblock Arithmetic in complex bases.
\newblock {\em Mathematics Magazine}, 57(2):77--81, 1984.

\bibitem[Gil96]{gilbert1996division}
William~J. Gilbert.
\newblock The division algorithm in complex bases.
\newblock {\em Canadian Mathematical Bulletin}, 39(1):47--54, 1996.

\bibitem[GKZ07]{gaudry2007gmp}
Pierrick Gaudry, Alexander Kruppa, and Paul Zimmermann.
\newblock A {GMP}-based implementation of {S}ch{\"o}nhage--{S}trassen's large
  integer multiplication algorithm.
\newblock In {\em Proceedings of the 2007 international symposium on Symbolic
  and algebraic computation}, pages 167--174, 2007.

\bibitem[GL18]{gu2018division}
Zhen Gu and Shuguo Li.
\newblock {{A division-free {T}oom--{C}ook multiplication-based montgomery
  modular multiplication}}.
\newblock {\em IEEE Transactions on Circuits and Systems II: Express Briefs},
  66(8):1401--1405, 2018.

\bibitem[GL20]{gu2020novel}
Zhen Gu and Shuguo Li.
\newblock A novel method of modular multiplication based on {K}aratsuba-like
  multiplication.
\newblock In {\em 2020 IEEE 27th Symposium on Computer Arithmetic (ARITH)},
  pages 33--40. IEEE, 2020.

\bibitem[GL21]{gu2021optimized}
Zhen Gu and Shuguo Li.
\newblock Optimized interpolation of four-term {K}aratsuba multiplication and a
  method of avoiding negative multiplicands.
\newblock {\em IEEE Transactions on Circuits and Systems I: Regular Papers},
  69(3):1199--1209, 2021.

\bibitem[GLPN93]{gallo-hypergraphs}
Giorgio Gallo, Giustino Longo, Stefano Pallottino, and Sang Nguyen.
\newblock Directed hypergraphs and applications.
\newblock {\em Discrete Applied Mathematics}, 42(2):177--201, 1993.

\bibitem[GM94]{granlund1994division}
Torbj{\"o}rn Granlund and Peter~L. Montgomery.
\newblock Division by invariant integers using multiplication.
\newblock In {\em Proceedings of the ACM SIGPLAN 1994 conference on Programming
  language design and implementation}, pages 61--72, 1994.

\bibitem[{GMP}23]{gmp}
{GMP development team}.
\newblock {\em {GNU MP}: the {GNU} multiple precision arithmetic library},
  6.3.0 edition, 2023.
\newblock \url{https://gmplib.org/}.

\bibitem[Hor19]{horner1819new}
William~George Horner.
\newblock A new method of solving numerical equations of all orders, by
  continuous approximation.
\newblock {\em Philosophical Transactions of the Royal Society of London},
  109:308--335, 1819.

\bibitem[HR92]{steiner-tree-problems}
Frank~K. Hwang and Dana~S. Richards.
\newblock Steiner tree problems.
\newblock {\em Networks}, 22(1):55--89, 1992.

\bibitem[HVDH21]{harvey2021integer}
David Harvey and Joris Van Der~Hoeven.
\newblock {{Integer multiplication in time {$O(n \log n)$}}}.
\newblock {\em Annals of Mathematics}, 193(2):563--617, 2021.

\bibitem[HVDHL16]{harvey2016even}
David Harvey, Joris Van Der~Hoeven, and Gr{\'e}goire Lecerf.
\newblock Even faster integer multiplication.
\newblock {\em Journal of Complexity}, 36:1--30, 2016.

\bibitem[HWDW23]{hao2023efficient}
Yue Hao, Weijiang Wang, Hua Dang, and Guiyu Wang.
\newblock Efficient {B}arrett modular multiplication based on {T}oom--{C}ook
  multiplication.
\newblock {\em IEEE Transactions on Circuits and Systems II: Express Briefs},
  2023.

\bibitem[IN89]{italiano1989online}
Giuseppe~F. Italiano and Umberto Nanni.
\newblock Online maintenance of minimal directed hypergraphs.
\newblock {\em Technical report, Columbia University}, 1989.

\bibitem[Jeb93]{jebelean-exact-div-93}
Tudor Jebelean.
\newblock An algorithm for exact division.
\newblock {\em Journal of Symbolic Computation}, 15(2):169--180, 1993.

\bibitem[JF06]{johnson2006modified}
Steven~G. Johnson and Matteo Frigo.
\newblock A modified split-radix {FFT} with fewer arithmetic operations.
\newblock {\em IEEE Transactions on Signal Processing}, 55(1):111--119, 2006.

\bibitem[KGNK23]{kalvacherla2023approximate}
Abhinav Kalvacherla, Rachana George, S.~Nalesh, and S.~Kala.
\newblock Approximate {CNN} on {FPGA} using {T}oom--{C}ook multiplier.
\newblock In {\em 2023 IEEE International Symposium on Smart Electronic Systems
  (iSES)}, pages 271--276. IEEE, 2023.

\bibitem[KKKH22]{kim2022toffoli}
Sunyeop Kim, Insung Kim, Seonggyeom Kim, and Seokhie Hong.
\newblock Toffoli gate count optimized space-efficient quantum circuit for
  binary field multiplication.
\newblock {\em Cryptology ePrint Archive}, 2022.

\bibitem[Knu81]{knuth1981semi}
D.~Knuth.
\newblock Seminumerical algorithms, 1981.

\bibitem[KO62]{karatsuba-multiplication}
Anatoly~A. Karatsuba and Yu.~P. Ofman.
\newblock Multiplication of many-digital numbers by automatic computers.
\newblock {\em Dokl. Akad. Nauk SSSR}, 145(2):293--294, 1962.

\bibitem[Kov16]{koval2016algorithm}
Aleksey Koval.
\newblock Algorithm for {G}aussian integer exponentiation.
\newblock In {\em Information Technology: New Generations: 13th International
  Conference on Information Technology}, pages 1075--1085. Springer, 2016.

\bibitem[KS15]{kepley2015quantum}
Shane Kepley and Rainer Steinwandt.
\newblock {{Quantum circuits for {$\mathbf{F}_2^{n}$}-multiplication with
  subquadratic gate count}}.
\newblock {\em Quantum Information Processing}, 14:2373--2386, 2015.

\bibitem[KS20]{karstadt}
Elaye Karstadt and Oded Schwartz.
\newblock Matrix multiplication, a little faster.
\newblock {\em Journal of the ACM (JACM)}, 67(1):1--31, 2020.

\bibitem[LAJK21]{larasati2021quantum}
Harashta~Tatimma Larasati, Asep~Muhamad Awaludin, Janghyun Ji, and Howon Kim.
\newblock Quantum circuit design of {T}oom 3-way multiplication.
\newblock {\em Applied Sciences}, 11(9):3752, 2021.

\bibitem[LM15]{lee2015subquadratic}
Chiou-Yng Lee and Pramod~Kumar Meher.
\newblock Subquadratic space-complexity digit-serial multipliers over {${\rm
  GF}(2^{m})$} using generalized (a,b)-way {K}aratsuba algorithm.
\newblock {\em IEEE Transactions on Circuits and Systems I: Regular Papers},
  62(4):1091--1098, 2015.

\bibitem[Low76]{Lowerre1976Harpy}
Bruce~T. Lowerre.
\newblock {\em The harpy speech recognition system.}
\newblock PhD thesis, USA, 1976.
\newblock AAI7619331.

\bibitem[MB19]{GaussianWinograd}
Lingchuan Meng and John Brothers.
\newblock Efficient {W}inograd convolution via integer arithmetic.
\newblock {\em arXiv preprint arXiv:1901.01965}, 2019.

\bibitem[MM60]{muir-treatise-determinants}
Thomas Muir and William~H. Metzler.
\newblock {\em A treatise on the theory of determinants}.
\newblock Dover Publications, Inc., 1960.

\bibitem[Mon85]{montgomery1985modular}
Peter~L. Montgomery.
\newblock Modular multiplication without trial division.
\newblock {\em Mathematics of computation}, 44(170):519--521, 1985.

\bibitem[Mor71]{morgenstern1971linear}
Jacques Morgenstern.
\newblock On linear algorithms theory of machines and computations.
\newblock {\em Academic Press (AP)}, 1971.

\bibitem[Mor73]{morgenstern1973note}
Jacques Morgenstern.
\newblock Note on a lower bound on the linear complexity of the fast {F}ourier
  transform.
\newblock {\em Journal of the ACM (JACM)}, 20(2):305--306, 1973.

\bibitem[Neu99]{Neukirch-Algebraic-NT}
J{\"u}rgen Neukirch.
\newblock {\em Algebraic number theory. {Transl}. from the {German} by
  {Norbert} {Schappacher}}, volume 322 of {\em Grundlehren Math. Wiss.}
\newblock Berlin: Springer, 1999.

\bibitem[NSS24a]{nissim2024fault}
Roy Nissim, Oded Schwartz, and Yuval Spiizer.
\newblock Fault-tolerant parallel integer multiplication.
\newblock In {\em Proceedings of the 36th ACM Symposium on Parallelism in
  Algorithms and Architectures}, pages 207--218, 2024.

\bibitem[NSS24b]{nissim2024minimizing}
Roy Nissim, Oded Schwartz, and Yuval Spiizer.
\newblock Minimizing {I/O} in {T}oom--{C}ook algorithms.
\newblock In {\em European Conference on Parallel Processing}, pages 59--73.
  Springer, 2024.

\bibitem[NSS25]{nissim2025minimizing}
Roy Nissim, Oded Schwartz, and Yuval Spiizer.
\newblock Minimizing the division factor of {T}oom--{C}ook algorithms.
\newblock In {\em 2025 Proceedings of the Conference on Applied and
  Computational Discrete Algorithms (ACDA)}, pages 144--154. SIAM, 2025.

\bibitem[NSS26]{Toom-Cook-divisions}
Roy Nissim, Oded Schwartz, and Yuval Spiizer.
\newblock The division barrier: optimal bounds and structural limits in
  {T}oom--{C}ook interpolation.
\newblock In {\em 2026 Annual ACM-SIAM Symposium on Discrete Algorithms
  (SODA)}. SIAM, 2026.

\bibitem[{Ope}26]{OAI:Integer-multiplication-below-n-log-n-September-23-2026}
{OpenAI}.
\newblock {Integer multiplication below $n\log n$}.
\newblock OpenAI Math Release preprint
  \href{https://github.com/openai/math/blob/main/preprints/Integer-multiplication-below-n-log-n-September-23-2026/paper.pdf}{OAI:Integer-multiplication-below-n-log-n-September-23-2026},
  2026.

\bibitem[Poh70]{Poh70}
Ira Pohl.
\newblock Heuristic search viewed as path finding in a graph.
\newblock {\em Artificial Intelligence}, 1(3-4):193--204, 1970.

\bibitem[PP87]{purdy-linear-division}
Carla~N. Purdy and George~B. Purdy.
\newblock Integer division in linear time with bounded fan-in.
\newblock {\em IEEE Transactions on Computers}, C-36(5):640--644, 1987.

\bibitem[PWLK23]{putranto2023space}
Dedy Septono~Catur Putranto, Rini~Wisnu Wardhani, Harashta~Tatimma Larasati,
  and Howon Kim.
\newblock Space- and time-efficient quantum multiplier in post quantum
  cryptography era.
\newblock {\em IEEE Access}, 11:21848--21862, 2023.

\bibitem[{Sag}22]{sagemath}
{Sage Developers}.
\newblock {\em {S}ageMath, the {S}age {M}athematics {S}oftware {S}ystem
  ({V}ersion 9.7)}, 2022.
\newblock \url{ https://www.sagemath.org}.

\bibitem[Set73]{sethi1973complete}
Ravi Sethi.
\newblock Complete register allocation problems.
\newblock In {\em Proceedings of the fifth annual ACM symposium on Theory of
  computing}, pages 182--195, 1973.

\bibitem[SKRK20]{srikanth2020enhancement}
B.~Srikanth, M.~Siva Kumar, J.~V.~R. Ravindra, and K.~Hari Kishore.
\newblock {{The enhancement of security measures in advanced encryption
  standard using double precision floating point multiplication model}}.
\newblock {\em Transactions on Emerging Telecommunications Technologies},
  31(12):e3948, 2020.

\bibitem[Spi22]{spiizer2022toom}
Yuval Spiizer.
\newblock Toom--{C}ook with base change.
\newblock Master's thesis, The Hebrew University of Jerusalem, 2022.
\newblock Available at:
  \url{http://www.cs.huji.ac.il/~odedsc/papers/Thesis-Yuval-Spiizer.pdf}.

\bibitem[SS71]{schonhage}
Arnold Sch{\"o}nhage and Volker Strassen.
\newblock Schnelle {M}ultiplikation grosser {Z}ahlen.
\newblock {\em Computing}, 7(3):281--292, 1971.

\bibitem[SS26]{libgtc2026}
Yuval Spiizer and Raphael~S. Steiner.
\newblock libgtc: long integer multiplication through {G}aussian integers, on
  top of {GMP}.
\newblock \url{https://github.com/huawei-csl/gaussian-toom-cook}, 2026.

\bibitem[SSSWD00]{SSSWD00}
Gerhard Schrimpf, Johannes Schneider, Hermann Stamm-Wilbrandt, and Gunter
  Dueck.
\newblock Record breaking optimization results using the ruin and recreate
  principle.
\newblock {\em Journal of Computational Physics}, 159(2):139--171, 2000.

\bibitem[Ste20]{destefani2020communicationoptimalparallelstandardkaratsuba}
Lorenzo~De Stefani.
\newblock Communication-optimal parallel standard and {K}aratsuba integer
  multiplication in the distributed memory model.
\newblock \url{https://arxiv.org/abs/2009.14590}, 2020.

\bibitem[TDR11]{TDR11}
Jordan~T. Thayer, Austin~J. Dionne, and Wheeler Ruml.
\newblock Learning inadmissible heuristics during search.
\newblock In {\em Proceedings of the 21st International Conference on Automated
  Planning and Scheduling (ICAPS 2011)}, pages 250--257, 2011.

\bibitem[THMK22]{thirumoorthi2022optimized}
Madhan Thirumoorthi, Moslem Heidarpur, Mitra Mirhassani, and Mohammed Khalid.
\newblock An optimized {$M$}-term {K}aratsuba-like binary polynomial multiplier
  for finite field arithmetic.
\newblock {\em IEEE Transactions on Very Large Scale Integration (VLSI)
  Systems}, 30(5):603--614, 2022.

\bibitem[Too63]{toom}
Andrei~L. Toom.
\newblock The complexity of a scheme of functional elements realizing the
  multiplication of integers.
\newblock In {\em Soviet Mathematics Doklady}, volume~4, pages 714--716, 1963.

\bibitem[Win71]{winograd1971multiplication}
Shmuel Winograd.
\newblock {{On multiplication of $2\times 2$ matrices}}.
\newblock {\em Linear algebra and its applications}, 4(4):381--388, 1971.

\bibitem[Win80]{winograd1980arithmetic}
Shmuel Winograd.
\newblock {\em Arithmetic complexity of computations}, volume~33.
\newblock Siam, 1980.

\bibitem[WPK24]{wardhani2024high}
Rini~Wisnu Wardhani, Dedy Septono~Catur Putranto, and Howon Kim.
\newblock High- and half-degree quantum multiplication for post-quantum
  security evaluation.
\newblock {\em IEEE Access}, 2024.

\bibitem[WYZ{\etalchar{+}}23]{wang2023tcpm}
Jianfei Wang, Chen Yang, Fahong Zhang, Yishuo Meng, and Yang Su.
\newblock {TCPM}: a reconfigurable and efficient {T}oom--{C}ook-based
  polynomial multiplier over rings using a novel compressed postprocessing
  algorithm.
\newblock {\em IEEE Transactions on Very Large Scale Integration (VLSI)
  Systems}, 31(8):1153--1166, 2023.

\bibitem[YS21]{yang2021521}
Hyeon-Jun Yang and Kyung-Wook Shin.
\newblock A $521$-bit high-performance modular multiplier using $3$-way
  {T}oom--{C}ook multiplication and fast reduction algorithm.
\newblock {\em Journal of the Korea Institute of Information and Communication
  Engineering}, 25(12):1882--1889, 2021.

\bibitem[Zan09]{zanoni2009Toom}
Alberto Zanoni.
\newblock {T}oom--{C}ook 8-way for long integers multiplication.
\newblock In {\em 2009 11th International Symposium on Symbolic and Numeric
  Algorithms for Scientific Computing}, pages 54--57. IEEE, 2009.

\bibitem[Zan10]{zanoni2010iterative}
Alberto Zanoni.
\newblock {{Iterative {T}oom--{C}ook methods for very unbalanced long integer
  multiplication}}.
\newblock In {\em Proceedings of the 2010 International Symposium on Symbolic
  and Algebraic Computation}, pages 319--323, 2010.

\bibitem[Zur93]{zuras1993squaring}
Dan Zuras.
\newblock On squaring and multiplying large integers.
\newblock In {\em Proceedings of IEEE 11th Symposium on Computer Arithmetic},
  pages 260--271. IEEE, 1993.

\end{thebibliography}
	\bibliographystyle{alpha}
	
\end{document}